%% file: ex_article.tex
\PassOptionsToPackage{dvipsnames}{xcolor} %Adding Color names - TC
\documentclass[onefignum,onetabnum]{siamart171218}

\input{ex_shared}

\ifpdf
\hypersetup{
  pdftitle={Exact and approximate solutions for elastic interactions in a nematic liquid crystal},
  pdfauthor={Thomas G.~J.~Chandler and Saverio E.~Spagnolie}
}
\fi

\begin{document}

\maketitle

% REQUIRED
\begin{abstract}
Anisotropic fluids appear in a diverse array of systems, from liquid-crystal displays to bacterial swarms, and are characterized by orientational order. Large colloidal particles immersed in such environments disturb the medium's orientational order,  resulting in a stored elastic energy within the bulk. As a consequence, multiple immersed bodies interact  at equilibrium through fluid-mediated forces and torques, which depend on the bodies' positions, orientations, and shapes. We provide the equilibrium configuration of a model nematic liquid crystal with multiple immersed bodies or inclusions in two-dimensions, as well as the associated body forces, torques, and surface tractions. A complex variables approach is taken which leans on previous work by Crowdy~\cite{Crowdy20} for describing solutions with multiply-connected domains. Free periods of a complex director field, which correspond to topological defect positioning and net topological charge, are determined numerically to minimize a global stored elastic energy, including a contribution of a weak (finite) anchoring strength on the body surfaces. Finally, a general, analytical description of two-body far-field interactions is provided, along with examples using two cylindrical inclusions of arbitrary position and size, and two triangles of arbitrary position and orientation.
\end{abstract}

% REQUIRED

% REQUIRED

\section{Introduction}\label{sec:intro}
Many fluids are host to a suspension of elongated bodies which show a preference towards orientational alignment. In a uniaxial liquid crystal (LC), the  local molecular orientation, averaged over a small control volume, is represented as a director field $\bm{n}(\bm{x},t)$, with  spatial position $\bm{x}$, time $t$, and $|\bm{n}(\bm{x},t)|=1$. Deformations of the director field away from uniformity result in an elastic stress response \cite{degennes1993,stewart2004}. Such fluids have been of great industrial interest for decades due to their optical properties \cite{yg09}, applicability to medical science \cite{wjc07}, chemical and biological sensing \cite{chmamta13}, and the design of soft active materials \cite{bbmwa16}. Active biological systems have been similarly described \cite{mjrlprs13,diys18}, from the dynamic ordering of mucus \cite{vhv93}, biofilms \cite{vcht08} and tissues \cite{sxll18,mnswsdmbkbk19} to suspensions of swimming bacteria \cite{ss08,ks11,ss15} and the interior of cells \cite{bn14,dejs17,nd17}. 

Among the most alluring (and analytically challenging) features of liquid crystals are the prevalence of topological singularities, which satisfy global conservation laws \cite{lpcs98,acmk12}. The locations of the defects on the surface or in the fluid depend on the relationship between the bulk elastic energy and the surface anchoring conditions on any domain boundaries. In addition to focusing elastic stress on immersed surfaces, topological defects are important sites in biological settings for the onset of cell death and extrusion \cite{sdnkttmltl17}, layer formation \cite{caws21}, cell accumulation \cite{kks17}, cell sorting \cite{bdsnmdtgsy21}, and morphogenesis \cite{mgslbk21,vm22,wmb23}. They have also been considered for directed self-assembly \cite{mstrz06,wmbda16} and control \cite{ptgwl16,gsla17,lbbss18,lybss19,fgsa22}. Analytical insight into defect positioning and its consequences for locally stored elastic energy is, thus, of broad interest. 

Bodies immersed in a LC (that are much larger than the LC constituents) disturb the orientational order of the bulk liquid crystal. Confining or immersed boundaries introduce preferential orientations of the director field  with a given strength (for instance, a tangential anchoring condition), these generally lie in competition with the preferred uniformity of the orientation field \cite{stark2001physics}. If there are multiple immersed bodies or boundaries, the elastic energy may be reduced by altering their relative positions and orientations. Dipolar and quadrupolar far-field interactions between colloids (depending on normal or tangential anchoring conditions) have been investigated in three-dimensions \cite{rnrp96,rt97,fsyy04,ablv23}, and similarly between a colloid and a confining boundary \cite{fly03}. When many colloids are introduced to a LC they can self-assemble into linear chains \cite{pslw97, lbp00,srztpbom07,dcd19}. When the bodies are sufficiently well separated, their long-range interactions conjure a related problem, the interaction of topological defects themselves \cite{ts17,hs20}.

Near-field interactions, meanwhile, can be strongly nonlinear due to the interaction and positional rearrangement of topological defects \cite{tspt02,atpad03,ca08,hs20}. The self-assembly of colloids in LCs has seen wide use in the engineering of smart materials, with applications ranging from biosensors to dynamic porous membranes \cite{bl21}. Rather than colloid translations and rotations to reduce the system energy, a separate path towards relaxation is available if the immersed particles are deformable \cite{md13,mpwsa16,zzmhap16,nesa20,sv22}.

As a consequence of defect repositioning in near-field interactions, spherical colloids with tangential anchoring can settle into a configuration with broken symmetry, and multiple colloids can self-assemble into a chain aligned at an angle of 30$^\circ$ with the alignment axis of the liquid crystal \cite{pw98,pfm99,slkkp05,tst12,darh18} or into kinked chains \cite{srztpbom08,gtvz23}. Crystal lattice configurations have also been observed \cite{nnl01,mstrz06,gtvz23}. More exotic interactions include particle binding via Saturn-ring defect interactions \cite{gkgad03, trahd12, tm13,Smalyukh20}. In addition to their positioning relative to the director field alignment axis, colloid interactions through the LC also depend on the particle geometry and relative orientation. Two triangular bodies, for instance, can be arranged such that they are either attractive or repulsive just by rotating them relative to one another \cite{lms09,Smalyukh20}.

While a variety of numerical methods for exploring LC configurations have been developed \cite{wzz21}, analytical solutions of the equilibrium director field configuration are needed in order to better understand the geometry-dependent, LC-mediated elastic body interactions. Even though the equilibrium director field is a harmonic function in the single Frank elastic constant approximation \cite{degennes1993}, these body interactions are not simple to determine due to nonlinear anchoring boundary conditions and topological defects, whose positions are unknown \emph{a priori}.

The equilibrium director field around a single immersed body already introduces a number of important features, which inform the question of body interactions. In Chandler and Spagnolie~\cite{cs23}, we used complex variables techniques to find analytical solutions in the asymptotic regime of large surface anchoring strengths. Among our findings, we showed that topological singularities are preferentially positioned at or near sharp corners of an immersed body, depending on whether the anchoring strength is infinite or finite. When multiple bodies are immersed in the fluid, or if a nearby wall or other boundary is present, the problem tends that much further from tractability. The complex variables approach for interactions was used to similarly characterize the interactions of two topological defects \cite{ts17}.

The problem of determining harmonic functions with generic boundary conditions in multiply-connected domains has been explored in great depth by Crowdy~\cite{Crowdy20}. Using complex variables, the problem can be recast as a search for a locally holomorphic function with particular boundary conditions. The physical domain is first conformally mapped to a multi-connected annulus, and then a series of images of a free-space Green's function across all of the (now circular) boundaries leads in the direction of the solution; although, additional care must be taken to monitor the periods of the holomorphic function and related auxillary functions around each boundary. 

\begin{figure}[ht!]
    \centering
    \includegraphics[width=.9\textwidth]{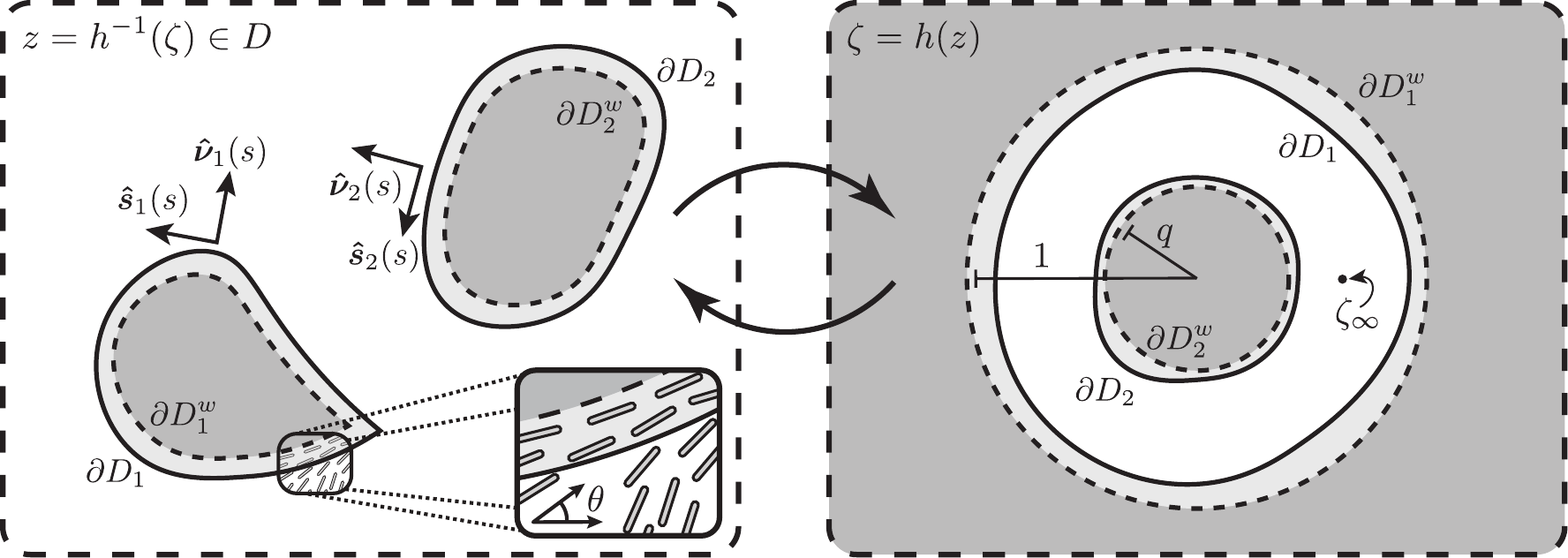}
    \caption{Left: the physical $z$-domain with two rigid bodies immersed in a two-dimensional nematic liquid crystal, where $z=x+i y$. The liquid crystal is described by a director field $\bm{n} = (\cos\theta, \sin\theta, 0)$ with director angle $\theta(z)\in[0,\pi)$ for $z\in D$. The boundaries of the two bodies are shown as solid curves, $\partial D_1$ and $\partial D_2$, with unit normal and tangent vectors $\bm{\n}_\ind(s)$ and $\bm{\s}_\ind(s)$, respectively. The effective (or virtual) boundaries are shown as dotted curves, $\partial D_1^w$ and $\partial D_2^w$. Right: the conformally-mapped $\zeta$-domain. The pole at $\zeta_\infty$ corresponds to $z\to \infty$ in the physical domain.} 
    \label{fig:sketch}
\end{figure}

In this paper, we use the approach put forth by Crowdy~\cite{Crowdy20} to analyze multibody interactions in a nematic liquid crystal. Figure~\ref{fig:sketch} provides a schematic of the general problem. Just as in the case of a single immersed body, the nonlinear boundary conditions and topological defect positions in the strong anchoring limit pose additional challenges, which are overcome upon appeals to the energy. Although a single body has no force or torque acting upon it \cite{cs23}, analogous to d'Alembert's paradox in classical potential flow theory, two bodies can impose forces and torques on each other through the fluid, as has been observed experimentally \cite{lms09}. These interactions are generically shape and orientation dependent. A number of additional nonlinear phenomena will be examined along the way, including a symmetry breaking instability when two cylinders are drawn nearer to each other, corresponding to a discontinuous jump in the topological defect positions.

This paper is organized as follows. We begin in \S\ref{sec:mathformulation} with a review of the mathematical model, including a discussion of boundary conditions and surface tractions, and we recall from Ref.~\cite{cs23} the effective boundary technique that allows for the solution of a weak (finite) anchoring problem based on the solution of a strong (infinite) anchoring problem with a slightly different boundary. Analytical solutions for two immersed bodies are then provided in \S\ref{sec:twoimmersedbodies}. Two worked examples of multiple-body interactions are then presented, which demonstrate the above methodology for determining the two-dimensional director field at equilibrium, including the selection of the topological charges and defect positions on the body surfaces. The first of these two examples is given in \S\ref{sec:ex1}, where we investigate two immersed cylinders with tangential anchoring, which includes the case of a single cylinder near an infinite wall as a limiting case.  We consider a more involved example in \S\ref{sec:ex2}, the interactions between two triangular prisms, where we again provide  formulae for the body forces and torques, and observe how defect positioning and particle interactions are orientation-dependent, reproducing experimental findings.  When the distance between the bodies is large, asymptotically valid approximations may be derived, as described in \S\ref{sec:asymptotics}. Finally, in \S\ref{sec:conc}, we  provide a closing summary and directions of future applications.

\section{Mathematical formulation}\label{sec:mathformulation}
We begin with a description of the general problem, and recall the relevant structure developed for the case of a single immersed body \cite{cs23}. Consider a two-dimensional nematic liquid crystal outside $N$ simply-connected bodies, as illustrated in Fig.~\ref{fig:sketch} for $N=2$. The liquid crystal domain and the boundary of the $\ind$th body are denoted by $D$ and $\partial \Di$, respectively. Assuming the one-constant approximation, the director angle, $\theta(x,y)$, is described by the Dirichlet free energy $\mathcal{F}_\mathrm{surface} \coloneqq K|\nabla\theta|^2/2$, where $K$ is the single Frank elastic constant. In general there are distinct elastic moduli penalizing LC bend and splay deformations, but they tend to be comparable \cite{brbf85,znnbsls14}, and the single constant model is often used to simplify mathematical analysis \cite{degennes1993}.

At the boundaries, the Rapini--Papoular form of the surface anchoring energy is given by $\mathcal{F}_\mathrm{surface} \coloneqq \Wi\sin^2(\theta-\phii)/2$, where $\Wi$ is the anchoring strength and $\phii$ is the preferred orientation defined on $\partial \Di$ \cite{rp69}. Examples will be provided for the important case where $\phii$ represents the tangent angle on the surface of the $\ind$th body, but the formulation below is valid for general $\phii$.

Combining the bulk and surface energies yields the net free energy
\begin{equation}\label{eq:totalenergy}
    \mathcal{E} \coloneqq \frac{K}{2}\iint_D|\nabla \theta|^2\de A +\sum_{\ind=1}^N  \frac{\Wi}{2}\int_{\partial \Di}\sin^2\left(\theta-\phii\right)\de s,
\end{equation}
where $s$ is an anti-clockwise arc length parameterization of the bodies, and $\de A$ and $\de s$ are the infinitesimal surface area and arc length elements, respectively. The principle of virtual work applied to \eqref{eq:totalenergy} yields the equilibrium equation for the director angle
\begin{equation}\label{eq:laplace}
    \nabla^2\theta = 0  \quad \text{in $D$},
\end{equation}
subject to the weak anchoring boundary conditions,
\begin{equation}\label{eq:weak_anchor}
-K\frac{\partial \theta}{\partial\n_\ind } + \frac{\Wi}{2}\sin\left[2(\theta-\phii)\right] =0 \quad \text{on $\partial \Di$},
\end{equation}
for $\ind\in\{1,\ldots N\}$, where $\bm{\n}_\ind=-\bm{x}_s^\perp$ is the fluid-pointing unit normal on the $\ind$th body, as depicted in Fig.~\ref{fig:sketch}. The traction on the $\ind$th surface due to the liquid crystal is also determined in this process (see Ref.~\cite{cs23}) and is given by
\begin{equation}\label{eq:traction}
    \bm{\bm{t}}_\ind  =  K\left(\frac{1}{2}|\nabla\theta|^2\bm{\n}_\ind -\frac{\partial \theta}{\partial\n_\ind } \nabla\theta\right)+ \frac{\Wi}{2}\Big(\sin(\theta-\phii)^2 \bm{\s}_\ind +\sin\left[2(\theta-\phii)\right]\bm{\n}_\ind  \Big)_{s},
\end{equation}
where $\bm{\s}_k =\bm{x}_s$ is the unit tangent vector on the $\ind$th body, and the subscript $s$ denotes an arc length derivative. Given a director field that satisfies \eqref{eq:laplace} and \eqref{eq:weak_anchor}, the energy and surface traction associated with the liquid crystal can be computed using \eqref{eq:totalenergy} and \eqref{eq:traction}, respectively. 

The problem is made dimensionless by scaling all lengths upon a characteristic length scale associated with the immersed bodies, $a$, and defining a dimensionless free energy, $\hatE\coloneqq \mathcal{E}/K$, and tractions, $\bm{\hat{t}}_{\ind}\coloneqq a^2 \bm{t}_{\ind}/K$. The resulting equations are governed by the dimensionless anchoring strengths $\wi \coloneqq a\Wi/K$. The dimensionless free energy of the liquid crystal may be written as a boundary integral using the divergence theorem, \ie
\begin{equation}\label{eq:totalenergy_dimensionless}
    \hatE = \frac{1}{2}\sum_{\ind=1}^N\int_{\partial \Di}-\theta \frac{\partial \theta}{\partial\n_\ind }+\wi\sin^2\left(\theta-\phii\right) \de s.
\end{equation}
Henceforth, we shall only work in these dimensionless variables.

\subsection{Complex variable representation}\label{sec:complex}
To access a wide range of complex variable techniques, we introduce the complex coordinate $z\coloneqq x+\im y$ and complex director angle 
\begin{equation}
    \Omega(z) \coloneqq \tau(x,y) - \im \theta(x,y),
\end{equation}
where $\tau(x,y)=\Re\Omega(z)$ is a harmonic conjugate of $\theta(x,y)=-\Im\Omega(z)$ (\ie~$\tau_x=-\theta_y$ and $\tau_y=\theta_x$) \cite{cs23}. Since $\theta(x,y)$ is harmonic in $D$, $\theta_x-\im\theta_y$ must be  holomorphic in $D$ and $\Omega(z)$ is at least locally-holomorphic. In general, $\Omega$ may not be single-valued around each immersed body, thus  the period  around  each must be defined. We write
\begin{equation}\label{eq:OmegaPeriod}
\oint_{\partial \Di}  \de \Omega \equiv \frac{1}{\im}\int_{\partial \Di}\theta_x -\im \theta_y \de z = \Ui-2\pi\im \Mi \qquad \text{for $\ind\in\{1,\ldots N$\}},
\end{equation}
for some given real constants $\Ui$ and half-integers $\Mi$, which correspond to  the topological charge of the $\ind$th body.

In these complex variables, the boundary condition \eqref{eq:weak_anchor} is equivalent to the constraints
\begin{equation}\label{eq:Omegabc}
\left(\big|\e^{\Omega(z)}\big|^2\right)_{s}+\wi\Im\left[ \e^{2\im\phii}\e^{2\Omega(z)}\right]=0 \quad \text{on $\partial \Di$},
\end{equation}
for $\ind\in\{1,\ldots,N\}$; the net free energy \eqref{eq:totalenergy_dimensionless} may be written as
\begin{equation}\label{eq:complexenergy}
\hatE = \frac{1}{4}\sum_{\ind=1}^N\oint_{\partial \Di}\Im\left[ \left(\Omega(z)- \overline{\Omega( z)}\right) \Omega'(z)z_{s}\right]+ \wi \Re\left[1-\e^{\Omega(z)-\overline{\Omega( z)}}\e^{2\im\phii} \right]\de {s},
\end{equation}
where the bar denotes a complex conjugate; and the surface traction on the $\ind$th body, $\bm{\hattau}_\ind\equiv(\hattau_\ind^x,\hattau_\ind^y)$ given by \eqref{eq:traction}, satisfies
\begin{equation}\label{eq:complextraction}
\hattau_\ind^x-\im \hattau_\ind^y=  \frac{1}{2\im}\Omega'(z)^2 z_{s} +\frac{\wi}{8}\left[\left(2+\e^{-2\im\phii}\e^{\overline{\Omega(z)}-\Omega(z)}-3\e^{2\im\phii}\e^{\Omega(z)-\overline{\Omega(z)}}\right)\bar z_{s}\right]_{s}.
\end{equation}
Integrating the traction around $\partial \Di$ yields the  net dimensionless force, $(\hatF_\ind ^x,\hatF_\ind ^y)$, and torque, $\hatT_\ind $, acting on the $\ind$th body:
\begin{subequations}\label{eq:complexforcetorque}
 \begin{gather}
 \hatF_\ind^x-\im \hatF_\ind^y  =\oint_{\partial \Di} \hat{t}_\ind^x-\im \hat{t}_\ind^y\de s = \frac{1}{2\im}\oint_{\partial \Di} \Omega'(z)^2 \de z, \\
 \hatT_\ind   =\oint_{\partial \Di}(x-x_\ind)\hat{t}_\ind^y -(y-y_\ind) \hat{t}_\ind^x\de s =\frac{1}{2}\Re\left[\oint_{\partial \Di} (z-z_\ind)\Omega'(z)^2 \de z\right]+\Ui,
\end{gather}
\end{subequations}
where $z_\ind=x_\ind+\im y_\ind$ is the centre of torque of the $\ind$th body.

\section{Analytical solutions for two immersed bodies}\label{sec:twoimmersedbodies}
In this section, we consider the interaction of two immersed bodies ($N=2$). The liquid crystal is assumed to be oriented with the $x$-axis in the far-field and subject to finite tangential anchoring on both $\partial D_1$ and $\partial D_2$, with dimensionless anchoring strengths $w_1$ and $w_2$, respectively;  that is, $\Omega(z)\to 0$ as $|z|\to\infty$ and  $\Omega(z)$ satisfies \eqref{eq:Omegabc} with $\phii=\arg( z_s)\mod\pi$.

We first show that the two bodies appear as equal and opposite topological charges in the far-field, and that their periods similarly sum to zero. Consider the contour integral $\oint_C\de \Omega$ for a  closed contour $C$ which encircles both bodies. Since $\Omega'(z)$ is  holomorphic in $D$, the contour can be deformed within $D$ via Cauchy's integral theorem. By taking the contour to infinity and imposing  the far-field condition, $\Omega(z)\to 0$ as $|z|\to \infty$, we find that the integral vanishes and, thus, $\Omega$ is single-valued outside the two bodies. However, $\Omega$ may be multi-valued along contours that pass between the  bodies. It follows that the period of $\Omega$ around $\partial D_1$ must be the additive inverse of the period around $\partial D_2$, \ie~$\Upsilon\coloneqq\Upsilon_1=-\Upsilon_2$ and $M\coloneqq M_1 =-M_2$ in \eqref{eq:OmegaPeriod}. The two bodies, thus, appear as topological defects of  charge $M_1=M$ and $M_2=-M$ in the far-field. We shall focus our attention on the case $M=0$ since this is known to minimize the free energy for an isolated body \cite{cs23}. 

At large anchoring strengths, subjecting a director field to finite-strength tangential anchoring on a boundary $\partial \Di$ (\ie~\eqref{eq:Omegabc} with $\phii=\arg (z_s)\mod\pi$) is asymptotically equivalent to subjecting it to strong (exact) tangential anchoring on an effective interior boundary $\partial \Di^w$, \ie
\begin{equation}\label{eq:strongtangential}
    \Im\left[\e^{\Omega(z)}z_s\right] =  \Oh(1/\wi^3)\quad \text{on $\partial \Di^\w$}
\end{equation}
as $\wi\to \infty$ \cite{cs23}. The effective (or virtual) boundary, $\partial \Di^w$, is found by displacing the physical boundary, $\partial \Di$, by  $-\bm{\n}_\ind  (s)/\wi-\bm{\s}_\ind'(s)/(2\wi^2) +\Oh(1/\wi^3)$. This asymptotic equivalence was termed the `effective boundary technique' \cite{cs23}.

Analytical progress can be made by writing the complex director as
\begin{equation}
    \Omega(z) = \log f'(z) + g(z),
\end{equation}
where $g(z)$ is any locally-holomorphic function that accounts for the periods in \eqref{eq:OmegaPeriod}, \ie
\begin{subequations}\label{eq:g_period}
\begin{equation}
    \oint_{\partial D_1} \de g  = \Upsilon \quad\text{and}\quad \oint_{\partial D_2} \de g = -\Upsilon, \subtag{a,b}
\end{equation}
\end{subequations}
and $f'(z)$ is a single-valued holomorphic function that accounts for the boundary conditions \eqref{eq:strongtangential}. Without loss of generality, we may choose $g(z)$ such that its imaginary part vanishes on  $\partial D_1^w$ and is constant on $\partial D_2^w$. The boundary conditions for $f(z)$ then follows from integrating \eqref{eq:strongtangential} with respect to arc length. Together these yield the  problem: find functions $f(z)$ and $g(z)$ such that
\begin{subequations}\label{eq:problem_g}
\begin{align}
g(z) \text{ locally holomorphic} &\qquad \text{ in $D^w$,}\\
 \Im g(z)=0 &\qquad \text{ on $\partial D_1^w$,}\\
  \Im g(z)=\alpha&\qquad \text{ on $\partial D_2^w$,}\\
 g(z)\to  \im \beta &\qquad \text{ as $|z|\to\infty$,}
\end{align}
\end{subequations}
with periods \eqref{eq:g_period} and real constants $\alpha$ and $\beta$, which are to be determined, and 
\begin{subequations}\label{eq:problem_f}
\begin{align}
    f(z) \text{ locally holomorphic} &\qquad \text{ in $D^w$,}\\
    \Im f(z)=0 &\qquad \text{ on $\partial D_1^w$,}\\
    \Im\left[e^{\im \alpha}f(z)\right]=C&\qquad \text{ on $\partial D_2^w$,}\\
    f(z)\sim   e^{-\im \beta}z &\qquad \text{ as $|z|\to\infty$,}
\end{align}
\end{subequations}
for some unknown constant $C$. Here, we have fixed the gauge of $f$  such that the constant in \myeqref{eq:problem_f}{b} vanishes. A unique solution is selected by specifying the period of $f$ around each body, \ie
\begin{equation}\label{eq:fCirc}
    \oint_{\partial D_1^w}  \de f =\Gamma_1 \qquad \text{and} \quad \oint_{\partial D_2^w}  e^{\im \alpha}\de f = \Gamma_2,
\end{equation}
for real periods $\Gamma_1$ and $\Gamma_2$.

By an extension of the Riemann mapping theorem for multi-connected domains (due to Koebe \cite{goluzin1969}), there exists a conformal map, $z=h(\zeta)$, from the doubly-connected  effective domain $z\in D^w$  to the  annulus $q\leq |\zeta|\leq 1$, with  $\partial D^w_1$ mapped onto $|\zeta|=1$, $\partial D^w_2$ onto $|\zeta|=q$, and $z=\infty$ to an interior point $\zeta=\zeta_\infty$ such that $z=h(\zeta)\sim C_\infty/(\zeta-\zeta_\infty)$ as $\zeta\to\zeta_\infty$. According to the Riemann mapping theorem, there are three real degrees of freedom for any conformal map of a simply-connected domain.  In the case of a doubly-connected domain, however, two of these degrees of freedom are needed to ensure the annulus is concentric. Thus, only a rotational degree of freedom remains, which we use to place $\zeta=\zeta_\infty$ on the positive real-axis, so that $0<q<\zeta_\infty<1$. The remaining   parameters $q$, $\zeta_\infty$, and $C_\infty$ are   dependent on the geometry and positions of the two bodies and must be determined.

In the $\zeta$-plane, the two potentials $G(\zeta)\coloneqq g(z(\zeta))$ and  $F(\zeta)\coloneqq f(z(\zeta))$ satisfy 
\begin{subequations}\label{eq:problem_G}
\begin{align}
G(\zeta) \text{ locally holomorphic} &\qquad \text{ in $q< \zeta< 1$,}\\
 \Im G(\zeta)=0 &\qquad \text{ on $|\zeta|=1$,}\\
  \Im G(\zeta)=\alpha&\qquad \text{ on $|\zeta|=q$,}\\
  G(\zeta)=\im  \beta  &\qquad \text{ at $\zeta = \zeta_\infty$},
\end{align}
\end{subequations}
with periods $\oint_{|\zeta|=1}\de G =-\Upsilon$ and $\oint_{|\zeta|=q}\de G = -\Upsilon$ and 
\begin{subequations}\label{eq:problem_F}
\begin{align}
F(\zeta) \text{ locally holomorphic} &\qquad \text{ in $q< \zeta< 1$,}\\
 \Im F(\zeta)=0 &\qquad \text{ on $|\zeta|=1$,}\\
  \Im\left[e^{\im \alpha} F(\zeta)\right]=C&\qquad \text{ on $|\zeta|=q$,}\\
F(\zeta)\sim C_\infty e^{-\im \beta} /(\zeta-\zeta_\infty) &\qquad \text{ as $\zeta\to \zeta_\infty$,}
\end{align}
\end{subequations}
with periods $\oint_{|\zeta|=1}\de F = -\Gamma_1$ and $ \oint_{|\zeta|=q} e^{\im \alpha}\de F = \Gamma_2$.

The analytical solution to \eqref{eq:problem_G} and \eqref{eq:problem_F} can be found  by using the method of images to construct functions akin to generalized Green's functions \cite{Crowdy20}. The full derivation is provided in Appendix~\ref{app:potential}. We find that
\begin{equation}\label{eq:potentialG_sol}
    G(\zeta) = -\frac{\Upsilon}{2\pi\im}\log\zeta,
\end{equation}
which yields the constants $\alpha = \Upsilon\log q/(2\pi)$ and $\beta=\Upsilon\log\zeta_\infty/(2\pi)$, and
\begin{equation}\label{eq:potentialF_sol}
\begin{split}
    F(\zeta)&= \frac{C_\infty e^{-\im\beta}}{\zeta_\infty}K(\zeta/\zeta_\infty) -\frac{\overline{C_\infty}e^{\im\beta}}{\zeta_\infty}K(\zeta_\infty\zeta)\\
    &\quad-\frac{\Gamma_1
    }{2\pi\im}\log\frac{P(\zeta/\zeta_\infty)}{P(\zeta_\infty\zeta)} - \frac{\Gamma_2 e^{-\im\alpha} }{2\pi\im}\log\frac{P(\zeta/\zeta_\infty)}{P(\zeta_\infty\zeta/q^2)},
    \end{split}
\end{equation}
where
\begin{subequations}\label{eq:PK}
\begin{align}
P(\zeta)&\coloneqq (1-\zeta)\prod_{k=1}^\infty (1-q^{2k}\zeta)^{e^{2k\im \alpha}}(1-q^{2k}/\zeta)^{e^{-2k\im \alpha}},\\
\text{and} \quad  K(\zeta)&\coloneqq\frac{\zeta P'(\zeta)}{P(\zeta)}=\frac{\zeta}{\zeta-1} +\sum_{k=1}^{\infty} \left(\frac{e^{-2k\im\alpha} q^{2k}}{\zeta-q^{2k}}-\frac{q^{2k}e^{2k\im\alpha}}{1/\zeta-q^{2k}}\right).  
\end{align}
\end{subequations}
Note that, since $0<q<\zeta_\infty<1$, this infinite product and summation  converge absolutely within the annulus, and  only a few terms are needed to obtain accurate approximations. The solution \eqref{eq:potentialF_sol} can be understood as follows: the first two terms account for the pole at $\zeta=\zeta_\infty$ by introducing an infinite cascade of images across the two body boundaries, $|\zeta| = 1$ and $|\zeta|=q$; and the final two terms account for the periods around the two bodies by introducing logarithmic cuts between $\zeta=\zeta_\infty$ and $\zeta=1/\zeta_\infty$ (\ie~across $|\zeta|=1$) and $\zeta=\zeta_\infty$ and $\zeta=q^2/\zeta_\infty$ (\ie~across $|\zeta|=q$). 

With these two potentials, the complex  director angle is 
\begin{equation}\label{eq:director_sol}
    \Omega(z) = \log f'(z)+g(z)= \log\left[F'(\zeta)/h'(\zeta)\right] -\Upsilon\log \zeta/(2\pi\im),
\end{equation}
where $\zeta = h^{-1}(z)$ and 
\begin{equation}\label{eq:dF}
\begin{split}
 F'(\zeta)&=C_\infty\frac{e^{-\im\beta}}{\zeta_\infty^2}K'(\zeta/\zeta_\infty)-\overline{C_\infty} e^{\im\beta}K'(\zeta_\infty\zeta)+\frac{\Gamma_2 e^{-\im \alpha}}{2\pi\im\zeta}\\
    &\quad+\frac{\Gamma_1+\Gamma_2 e^{\im \alpha}
    }{2\pi\im\zeta}K(\zeta_\infty\zeta)-\frac{\Gamma_1
    +\Gamma_2 e^{-\im \alpha}}{2\pi\im\zeta}K(\zeta/\zeta_\infty).
    \end{split}
\end{equation}
The above expression has been simplified using the identity $P(\zeta/q^2)= -\zeta P(\zeta)^{e^{2\im\alpha}}/q^2$. Finally, the director angle is given by $\theta = -\Im\Omega(z)$, and the  free energy, surface tractions,  body forces,  and body torques are computed by inserting \eqref{eq:director_sol} into \eqref{eq:complexenergy}--\eqref{eq:complexforcetorque}. Analyzing these solutions further, however, requires the specification of the physical domain (\ie~the conformal map $h(\zeta)$). We, therefore, proceed to consider two concrete examples: two interacting circular cylinders and  two interacting triangular prisms. Following these, we will provide a more general analytical perspective which is available when the bodies are well separated.

\section{Example 1: Two cylinders with tangential anchoring}\label{sec:ex1}
\begin{figure}[ht!]
    \centering
    \includegraphics[width=0.75\textwidth]{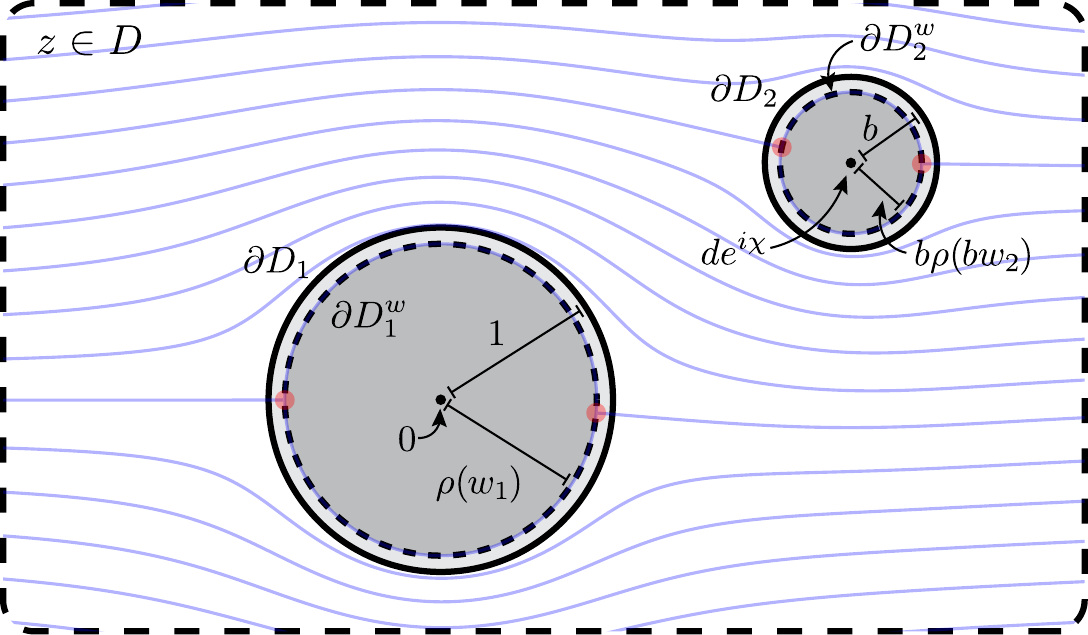}
    \caption{Example 1. Two-dimensional liquid crystal outside a unit cylinder centred at $z=0$ and a cylinder of radius $b$ centered at $z=d e^{\im\gamma}$ ($\partial D_1$ and $\partial D_2$, black solid curves). The  effective domain boundaries are shrunken cylinders of radii $\rho(w_1)$ and $b\rho(b w_2)$, respectively ($\partial D_1^w$ and $\partial D_2^w$, black dotted curves). Integral curves of the director field are shown  in blue for $w_1=w_2=10$, $d=2.75$, $b=0.5$, $\chi=\pi/6$, and using numerically determined energy-minimizing periods:  $\Gamma^{\min}_1\approx-0.3097$, $\Gamma^{\min}_2\approx 0.3786$, and $\Upsilon^{\min}=-0.0225$.}
    \label{fig:ex1_sketch}
\end{figure}

Consider a liquid crystal outside two immersed  cylinders, one centred at $z=0$ with dimensionless unit radius, and the other centred at $z=d e^{\im\chi}$ with dimensionless radius $b$. We denote the boundaries of these cylinders as $\partial D_1$ and $\partial D_2$, respectively. The liquid crystal is assumed to be oriented with the $x$-axis in the far-field and is subject to finite tangential anchoring on each cylinder: $\Omega(z)\to 0 $ as $|z|\to \infty$ and $\Omega(z)$ satisfies \eqref{eq:Omegabc} with $\phi=\arg (z_s)\mod\pi$ on the surfaces $|z|=1$ and $|z-d e^{\im\chi}|=b$. We shall also assume that both cylinders have vanishing topological charge, \ie~$\oint_{\partial \Di}\de\theta=0$. This configuration is plotted in Fig.~\ref{fig:ex1_sketch}.

For large anchoring strengths (\ie~$w_1\gg 1$  and $w_2\gg 1$), the effective boundary technique may be implemented and the solutions  derived in \S\ref{sec:twoimmersedbodies} may be used. The first step is to find the  effective boundaries corresponding to the two cylinders, $|z|=1$ and $|z-d e^{\im\chi}|=b$. In Ref.~\cite{cs23}, we showed that the effective boundary corresponding to a unit cylinder with anchoring strength $w$ is a cylinder of  radius $|z|=\rho(\w)\coloneqq(\sqrt{1+4/w^2 }-2/w)^{1/2}$. This effective boundary  is not only consistent with the asymptotic expression in \eqref{eq:strongtangential} for large $w$, but it in fact holds for all anchoring strengths (\ie~$w\geq 0$). It follows that, here, a suitable choice for the effective boundaries of the two cylinders is $|z|=\rho(\w_1)$ and $|z-d e^{\im\chi}|=b\rho(b \w_2)$, which we denote as $\partial D_1^w$ and $\partial D_2^w$, respectively. 

The next step is to find a conformal map, $z=h(\zeta)$, which maps the effective domain $z\in D^w$ to the annulus $q\leq|\zeta|\leq 1$. Consider the M\"obius transformation,
\begin{equation}\label{eq:ex1_mobius}
   z=h(\zeta)=\rho(w_1)e^{\im\chi}\,\frac{\zeta_\infty \zeta-1}{\zeta-\zeta_\infty}.
\end{equation}
This map is a composition of three conformal maps: $Z\coloneqq ze^{-\im\chi}/\rho(w_1)$  rotates and expands the domain so that the primary cylinder is of unit size and the secondary cylinder lies on the positive real axis; $\eta\coloneqq 1/Z$ reflects the exterior of the unit cylinder into the interior; and $\eta \coloneqq (\zeta-\zeta_\infty)/(\zeta_\infty\zeta-1)$ is an automorphism of the unit disc, which centres the eccentric circles. Without loss of generality, we place $\zeta_\infty$ on the positive real axis. The resulting transformation, \eqref{eq:ex1_mobius}, maps  $z=\infty$ onto $\zeta= \zeta_\infty$ and $\partial D_w^1$ onto $|\zeta| = 1$, whilst $\zeta_\infty$ and $q$ are chosen  such that  $\partial D^w_2$ is mapped onto $|\zeta|=q$.

Using the results of \S\ref{sec:twoimmersedbodies}, the complex director angle, $\Omega=\tau-\im\theta$, is given by the expression \eqref{eq:director_sol} with $\zeta=h^{-1}(z)=(\zeta_\infty z- \rho(w_1)e^{\im\chi})/(z-\zeta_\infty\rho(w_1)e^{\im\chi})$ and $C_\infty=-(1-\zeta_\infty^2)\rho(w_1)e^{\im\chi}$. The three periods in \eqref{eq:director_sol}, \ie~$\Gamma_1$, $\Gamma_2$, and $\Upsilon$, still remain unknown. Determining these requires the computation and minimization of the free energy \eqref{eq:complexenergy}. Before we address this, however,  it is useful to analyze the singularities of $\Omega(z)$ (\ie~the topological defects) within the effective domain, $D^w$.

\subsection{Topological defects}\label{sec:cylinder_defects}
The  director field corresponding to \eqref{eq:director_sol} does not contain defects in the fluid domain since $\Omega(z)$ is analytic by construction. There are, however, two  $-1$  defects on the boundary of each effective cylinder, that is 
\begin{subequations}\label{eq:effective-surface-defects}
    \begin{align}
        \Omega(z)&\sim\log\left[z-\rho(w_1)e^{\im a_1^\pm}\right]  &&\text{as $z\to\rho(w_1)e^{\im a_1^\pm}$},\\
 \text{and}  \quad  \Omega(z)&\sim\log\left[z-de^{\im\chi}-b \rho(b w_2)e^{\im a_2^\pm}\right]  &&\text{as $z\to de^{\im\chi}+b \rho(b w_2)e^{\im a_2^\pm}$,}
\end{align}
\end{subequations}
for some real constants $a_1^\pm$ and $a_2^\pm$. These defects tend to points on the body surfaces in the limits as $\rho(w_1\to\infty)\to 1$ and $\rho(b w_2\to\infty)\to 1$. Furthermore, we will show that  their positions (\ie~the arguments $a_1^\pm$ and $a_2^\pm$) are of the utmost importance for understanding  body interactions within the liquid crystal, and  we shall refer to these as `effective-boundary defects'. (Note that there are in fact a countably infinite number of `defects' within the two cylinders, corresponding to the singularities of the analytical continuation of $\Omega(z)$ within $|z|\leq \rho(w_1)$ and $|z- d e^{\im\chi}|\leq b\rho(bw_2)$. These singularities are the images of the above four effective-boundary defects and  $\zeta=\zeta_\infty$ across $|\zeta|=1$ and $|\zeta|=q$ --- a consequence of the method of images.)

For  a single immersed body, in Ref.~\cite{cs23}, we showed that the positions of  two effective-boundary defects are set by a single unknown period $\Gamma$, which is determined by minimizing the free energy of the liquid crystal. Analogously, here the positions of the four effective-boundary defects are set by minimizing the energy for the three unknown periods, $\Gamma_1$, $\Gamma_2$, and $\Upsilon$. By inserting the Schwarz functions of the two circles, $\bar{z}=S_1(z) = 1/z$ for $\partial D_1$ and $\bar{z}= S_2(z)\coloneqq d e^{-\im\chi}+b^2/(z-de^{\im\chi})$ for $\partial D_2$, into the expression \eqref{eq:complexenergy} with $e^{\im\phii}=z_s = 1/\sqrt{S'_\ind (z)}$, the net free energy  appears as the real part of a sum of two  closed  contour integrals (corresponding to the two bodies). These integrals are computed numerically in order to determine the energy-minimizing values $\Gamma_1=\Gamma_1^{\min}$, $\Gamma_2=\Gamma_2^{\min}$, and $\Upsilon=\Upsilon^{\min}$, which we pursue using the Nelder--Mead simplex search method (\textsc{Matlab}'s \texttt{fminsearch}) \cite{lrww98}. Figure~\ref{fig:ex1_sketch} shows integral curves of the energy-minimizing director field in blue for the given physical configuration, whilst the effective-boundary defects are shown as red dots. A loss of symmetry due to the two bodies is apparent.

When the  two effective cylinders have equal radii (\ie~$\rho(w_1)=b\rho(b w_2)$, for example when $b=1$ and $w_1=w_2$), the domain is symmetric across the line $y\sin\chi= -x\cos \chi$ and one finds that $\zeta_\infty=\sqrt{q}$. % = d/(2\rho(w_1)) - \sqrt{d^2/(4\rho(w_1)^2)-1}$. 
After minimizing the energy, the positions of the effective-boundary defects are also found to be symmetrically located with $\Gamma_1^{\min}=-\Gamma_2^{\min}$ and $\Upsilon^{\min}=0$. The complex director angle, \eqref{eq:director_sol}, then takes the simplified form 
\begin{equation}
\begin{split}
    \Omega(z) &= \log\left(\sum_{k=-\infty}^{\infty} \left[\frac{ q^{2k}}{(\zeta-q^{2k}\zeta_\infty)^2}- \frac{ q^{2k}e^{-2\im\chi}}{(\zeta_\infty\zeta-q^{2k})^2} \right]+\frac{\im e^{-\im\chi}G^{\min} }{\zeta}\right)\\
    &\quad+2\log\left(\zeta-\zeta_\infty\right),
    \end{split}
\end{equation}
where $\zeta(z)=h^{-1}(z)=(\zeta_\infty z- \rho(w_1)e^{\im\chi})/(z-\zeta_\infty\rho(w_1)e^{\im\chi})$ and $\Gamma_1^{\min}=-\Gamma_2^{\min} = 2\pi(1-\zeta_\infty^2)\rho(w_1)G^{\min}$.

\subsection{Body forces and symmetry breaking configurations}
Changing the body positions results in a change in the total elastic energy stored in the fluid --- external forces are, thus, required to keep the bodies fixed in place. The force acting on a body is found by integrating $\Omega'(z)^2/(2\im)$  around a closed contour containing it, \ie~\myeqref{eq:complexforcetorque}{a}. Integrating around a closed contour containing both bodies yields the total force acting on the system, but since $\Omega(z)\sim 0$ as $|z|\to \infty$, this integral must vanish \cite{cs23}. It follows that  the force acting on one of the cylinders is equal and opposite to the force acting on the other cylinder, \ie~$(\hatF_2^x, \hatF_2^y)=-(\hatF_1^x, \hatF_1^y)$. We compute this force using adaptive quadrature in \textsc{Matlab}.

\begin{figure}[ht!]
    \centering
    \includegraphics[width=0.9\textwidth]{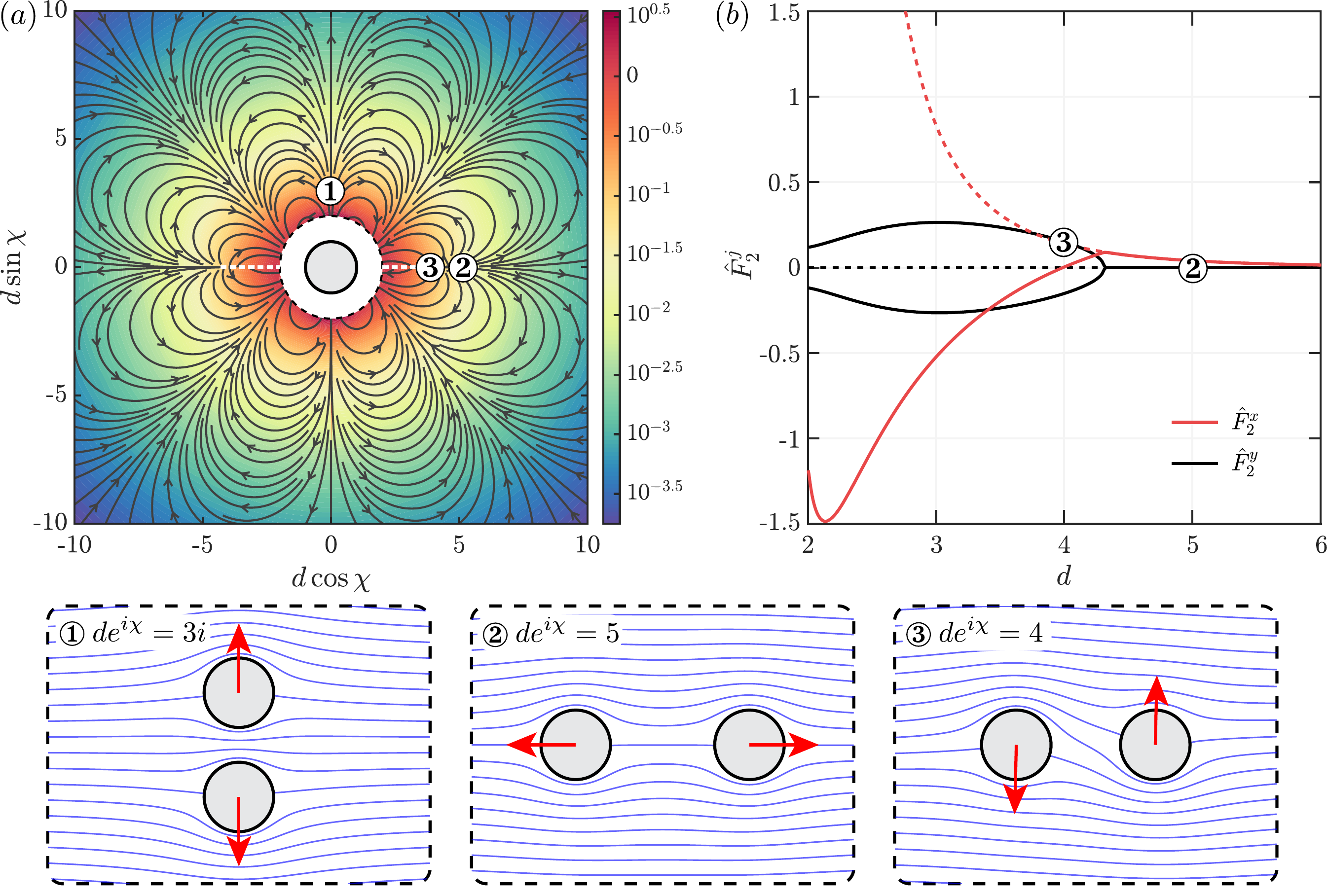}
    \caption{Example 1. (a) A contour plot showing the dimensionless net force, $(\hatF_x^2,\hatF_y^2)$, acting on a cylinder placed at $|z-d e^{\im\chi}|=b$ due to a unit cylinder  at $z=0$ for  $w_1=w_2=10$, $b=1$, and energy-minimizing periods $\Gamma_1^{\min}=-\Gamma_2^{\min}$ and $\Upsilon^{\min}= 0$. Arrows denote the direction of the force and colour denotes the magnitude, $|\hatF_x^2-\im \hatF_y^2|$. The cylinder cannot be placed inside $|z|=1+b=2$ (dashed curve) due to the unit cylinder at $z=0$ (solid curve).  (b) The dimensionless net force acting on inline cylinders ($\chi=0$) is shown as solid lines for the energy-minimizing periods and dashed lines for vanishing periods ($\Gamma_1=\Gamma_2=\Upsilon=0$). These solutions diverge as $d$ decreases, resulting in a supercritical pitchfork bifurcation. The multiple energy minimizing states is delineated by the white dashed lines in (a). In \ding{172}--\ding{174}, the integral curves of the director field are shown as blue curves for $de^{\im\chi}=3\im$, $5$, and $4$, whilst the direction of the force acting on the two cylinders is shown by the red arrows.}
    \label{fig:ex1_Force}
\end{figure}

The force acting on $\partial D_2$ is plotted in the phase portrait in Fig.~\myref{fig:ex1_Force}{a} for $b=1$, $w_1=w_2=10$, and the numerically determined energy-minimizing periods $\Upsilon^{\min}= 0$ and $\Gamma_1^{\min}=-\Gamma_2^{\min}$. Furthermore, three examples of the forces on the cylinders at different body configurations, as well as the quasi-static LC director field, are presented in Fig.~\myrefnb{fig:ex1_Force}{\ding{172}--\ding{174}}. If the line of centers between the particles is either parallel or anti-parallel to the alignment axis, the bodies experience a repulsion from one another, provided they are sufficiently separated (Fig.~\myrefnb{fig:ex1_Force}{\ding{172}--\ding{173}}). But this interaction is unstable to symmetry-breaking perturbations. For instance, if the angle between the body centers is small, but nonzero, body forces would seek to increase this angle. Additionally, when the cylinders are inline ($\chi=0$ or $\pi$), the energy-minimizing configuration undergoes a supercritical pitchfork bifurcation as the separation distance, $d$, decreases; this is delineated by the white dashed line in  Fig.~\myref{fig:ex1_Force}{a} and is shown explicitly in Fig.~\myref{fig:ex1_Force}{b}. This bifurcation is a result of the effective-boundary defects transitioning from being up--down symmetric  (Fig.~\myrefnb{fig:ex1_Force}{\ding{173}}) to being off-axis (Fig.~\myrefnb{fig:ex1_Force}{\ding{174}}). Setting $\Upsilon=\Gamma_1=\Gamma_2=0$, instead of minimizing the energy,  fixes the locations of the defects to always be up--down symmetric, resulting in the cylinders being repulsive at all separation distances --- see the dashed line in  Fig.~\myref{fig:ex1_Force}{b}. This symmetry breaking, and associated snapping from repulsion to attraction, has been observed for spheres with tangential anchoring as well \cite{slkkp05,tst12}.

If the bodies were free to move (and the relaxation time of the liquid crystal was sufficiently small so that a quasi-static approximation could be made), the cylinders would eventually attract each other along a path diagonal to the preferred orientation of the liquid crystal. Thus, while spheres with strong tangential anchoring have been found to align experimentally at a $30^\circ$ offset from the alignment axis \cite{pw98,pfm99,slkkp05}, cylinders are predicted to align at a $45^\circ$ offset. More generally, the offset angle of the stable configuration is dependent on the anchoring strengths and the ratio of cylinder radii, $b$. For example, when $w_1=w_2=10$ and $b=1$ (as pictured in Fig.~\ref{fig:ex1_Force}) the stable configurations are at a much smaller offset angle of approximately $8^\circ$. This is substantially smaller than the chaining angle observed in two dimensions for two sharp bodies \cite{darh18}, suggesting that corners, which promote defect repositioning, can have an outsized effect. Fixing the locations of the defects, instead of minimizing the energy for  their locations, results in a comparable offset angle.

At large separation distances, the  bodies only weakly interact and the force appears to resemble an asymptotic pole of the from
\begin{equation}\label{eq:symmetriccylinder_asymforce}
    \hatF_1^x-\im \hatF_1^x= -\hatF_2^x+\im \hatF_2^x\sim -C/\left(de^{\im\chi}\right)^5 \quad \text{as $d\to\infty$,}
\end{equation}
 for some $C>0$, which is consistent with previous work on far-field quadrupolar interactions in a nematic LC \cite{lcty02}. We will return to this topic for arbitrary particle shapes and sizes in \S\ref{sec:asymptotics}, but first we explore a second example.

\section{Example 2: Two triangles with tangential anchoring}\label{sec:ex2}
\begin{figure}[ht!]
    \centering
    \includegraphics[width=0.75\textwidth]{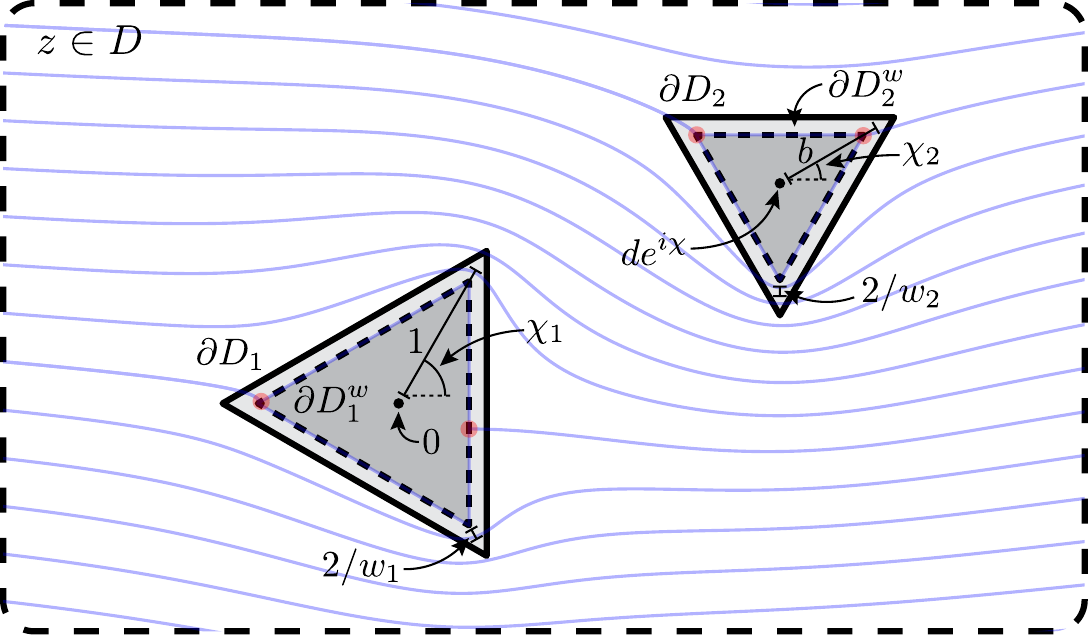}
    \caption{Example 2.  Two-dimensional liquid crystal outside two triangles with corners at the roots of $z^3=e^{3i\chi_1}$ and $(z-de^{\im\chi})^3=b^3e^{3i\chi_2}$ ($\partial D_1$ and $\partial D_2$, black solid lines). The  effective domain boundaries are similar triangles with corners at the roots of $z^3=(1-2/w)^3e^{3i\chi_1}$ and $(z-de^{\im\chi})^3=(b-2/w)^3 e^{3i\chi_2}$  ($\partial D_1^w$ and $\partial D_2^w$, black dotted lines). Integral curves of the director field are shown in blue for $w=10$, $d=2.5$, $b=0.75$, $\chi_1=\pi/3$, $\chi_2=\pi/6$, $\chi = \pi/6$, and energy-minimizing periods: $\Gamma^{\min}_1=0.00$, $\Gamma^{\min}_2=2.41$, and $\Upsilon^{\min}=-0.01$.}
    \label{fig:ex2_sketch}
\end{figure}

In this section, we consider two equilateral triangles, one with corners at the roots of $z^3 = e^{3\im\chi_1}$ and the other with corners at the roots of $(z-d e^{\im\chi})^3 = b^3e^{3\im\chi_2}$. We denote these triangles as $\partial D_1$ and $\partial D_2$, respectively. As before, the LC is assumed to be oriented with the $x$-axis in the far-field and subject to finite tangential anchoring on each triangle. Both triangles are also assumed to have vanishing topological charge, \ie~$\oint_{\partial \Di}\de\theta=0$. This configuration is plotted in Fig.~\ref{fig:ex2_sketch}. 

Here we again make use of the effective boundaries, internal to the physical boundaries, upon which the anchoring is strong (\ie~perfect tangential anchoring). Since angles are preserved under the effective boundary technique, the triangle $\partial D_1$ is mapped to a similar triangle with corners at the roots of  $z^3 = (1-2/w_1)^3e^{3\im\chi_1}$ \cite{cs23}, which we denote as $\partial D_1^w$. Likewise, the triangle $\partial D_2$ is mapped to a similar triangle with corners at the roots of $(z-d e^{\im\chi})^3 = (b-2/w_2)^3e^{3\im\chi_2}$, which we denote as $\partial D_2^w$. These effective triangles are shown in Fig.~\ref{fig:ex2_sketch} as dotted lines. 

Next, we seek a conformal map, $z=h(\zeta)$, that maps the effective domain $z\in D^w$ onto the annulus $q\leq|\zeta|\leq 1$. This is achieved by using an extension to the Schwarz--Christoffel mapping for multiply-connected polygonal domains \cite{Crowdy2005}. The mapping takes the form
\begin{equation}\label{eq:ex2_SchwarzChristofell}
    z = h(\zeta) = A+B\int^\zeta \frac{\prod_{k=1}^3  \left[P_0(s/a_1^k) P_0(s/a_2^k)\right]^{2/3}}{\left[sP_0(s/\zeta_\infty)P_0(s \bar{\zeta}_\infty)\right]^2}\de s,
\end{equation}
where $A$ and $B$ are complex constants and
\begin{equation}
P_0(\zeta)\coloneqq (1-\zeta)\prod_{k=1}^\infty (1-q^{2k}\zeta)(1-q^{2k}/\zeta),
\end{equation}
\ie~\myeqref{eq:PK}{a} with $\alpha=0$. Here, $\zeta=\zeta_\infty$ is the image of $z=\infty$ and $\zeta=a_1^k$ and $\zeta=a_2^k$ are the images of the corners of $\partial D_1^w$ and $\partial D_2^w$ on $|\zeta|=1$ and $|\zeta|=q$, respectively. Without loss of generality, we shall place  $\zeta=\zeta_\infty$ on the positive real-axis by setting the rotational degree of freedom of the annulus. The remaining twelve accessory parameters  ($A$, $B$, $q$, $\arg a_j^k$,  and $|\zeta_\infty|$) are determined by ensuring the six vertices are mapped correctly.
 
Determining the accessory parameters of a Schwarz--Christoffel mapping is itself a challenging problem \cite{dt02}, and we will turn to numerical techniques. The \textsc{Matlab} toolbox \texttt{plgcirmap} \cite{nasser2020} computes the conformal mapping from a given  multiply-connected polygonal domain onto a circular domain. We use this package to compute the mapping from the effective domain, $z\in D^w$, onto an auxiliary domain outside two circular cylinders, $|\eta|=r_1$ and $|\eta-D e^{\im X}|=r_2$, whilst preserving orientation at infinity, \ie~$z\sim \eta +\Oh(1/\eta)$ as $|\eta|\to \infty$. Here, $r_1$, $r_2$, $D$, and $X$ are numerically determined real numbers. We then apply a M\"obius transformation that maps this auxiliary domain onto the annulus $q\leq |\zeta|\leq 1$, that is
\begin{equation}
\eta(\zeta)  = r_1 e^{\im X}\frac{\zeta_\infty\zeta-1}{\zeta-\zeta_\infty},
\end{equation}
where $\zeta_\infty$ and $q$ are chosen such that $|\eta-D e^{\im X}|=r_1$ is mapped onto $|\zeta|=q$ --- this final mapping is analogous to  \eqref{eq:ex1_mobius}. The composition of these two conformal maps is equivalent to computing the Schwarz--Christoffel mapping \eqref{eq:ex2_SchwarzChristofell}, by the uniqueness of conformal mappings  \cite{ablowitz2003,goluzin1969}.

\subsection{Topological defects}\label{sec:triangles_defects}
Equipped with the above conformal map, the complex director angle, $\Omega(z)=\tau(x,y)-\im\theta(x,y)$, can be expressed as  \eqref{eq:director_sol} with  $C_\infty=-r_1 e^{\im X}(1-\zeta_\infty^2)$. By construction, $\Omega(z)$ is analytic outside $\partial D_1^w$ and $\partial D_2^w$, thus there are no topological defects in the fluid domain. However, defects appear at the corners of the effective triangles (a consequence of the Schwarz--Christoffel mapping), as well as two additional $-1$ defects on each effective boundary. These `effective-boundary defects' are akin to those found for the two immersed cylinders, \eqref{eq:effective-surface-defects}, and their positions are controlled by the three periods, $\Gamma_1$, $\Gamma_2$, and $\Upsilon$. As before, these periods are determined by numerical minimization of the free energy, \eqref{eq:complexenergy}. This is achieved by parameterizing each side of the triangle, and evaluating the integral and conformal map numerically for given  $b$, $d$, $\chi$, $\chi_1$, $\chi_2$, $w_1$, $w_2$, $\Gamma_1$, $\Gamma_2$, and $\Upsilon$. The resulting energy is then minimized to determine $\Gamma_1=\Gamma_1^{\min}$, $\Gamma_2=\Gamma_2^{\min}$, and $\Upsilon=\Upsilon^{\min}$. 

An example of a director field found in the manner above is shown as blue curves in Fig.~\ref{fig:ex2_sketch}, whilst the effective-boundary defects are shown as red dots. Note that three  defects are located at the corners of the effective boundaries, whilst the fourth sits along one straight edge. Corners are natural locations for topological defects to reside in order to reduce the total elastic energy in the LC, just as the Kutta condition selects the circulation (by the placement of a surface stagnation point at a sharp edge) in potential flow theory \cite{cs23}.  In general, the final effective-boundary defect is not found at a corner due to the constraint of horizontal alignment in the far-field. Thus, one triangle is allowed to have both defects at corners (with 3 possible pairs)  and the other triangle is  only allowed  one  defect at a corner (again, with 3 possible choices).   Each of the $3\times 3\times 2=18$ possible combinations of corners for the effective-boundary defects to be located correspond to local energy minimizing configurations. Determining precisely which three corners yield the global energy minimum requires direct comparison. The LC configuration around a single regular polygon has also been studied using a reduced Landau--de Gennes framework \cite{hmz20,hm23}.

\begin{figure}[ht!]
    \centering
    \includegraphics[width=0.9\textwidth]{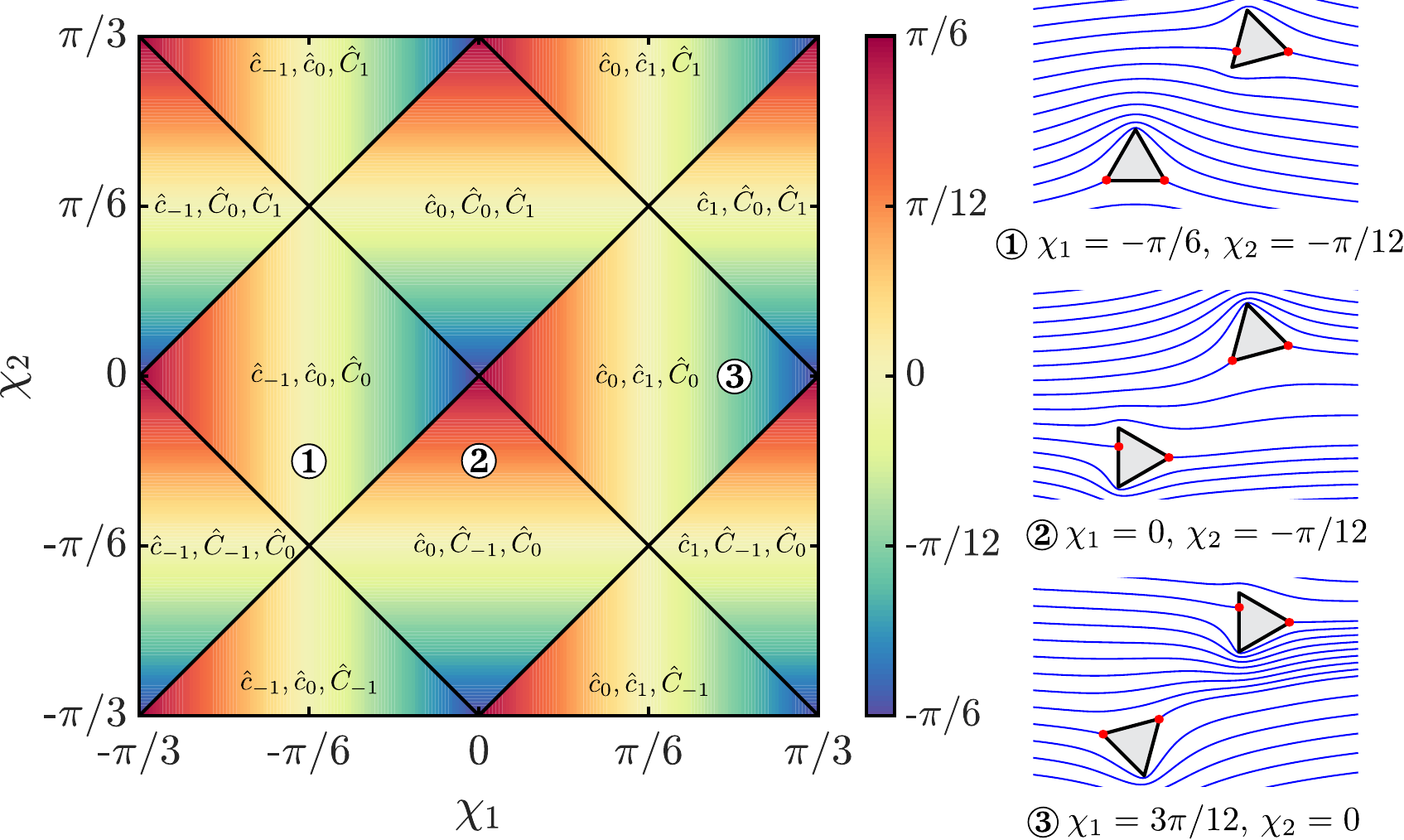}
    \caption{Example 2. Plot of the triangles' orientation  $(\chi_1,\chi_2)$-space showing the three  triangle corners at which an effective-boundary defect is located in the strong anchoring limit, $w_1,w_2\to \infty$. Here, the black solid lines partition the parameter space, whilst  $\hat{c}_{k}=(1-2/w_1)e^{\im\chi_1}e^{2\im\pi  k/3}$ and $\hat{C}_k= de^{\im\chi}+(b-2/w_2)e^{\im\chi_2}e^{2\im\pi  k/3}$ with  $k\in\{-1,0,1\}$ denote  the six corners of the effective triangles. Color is proportional to the asymptotic torques $\hat T_1\sim -\hat T_2$ as $d\to\infty$, as defined  in \eqref{eq:triangle_asymtorque}. These torques drive the triangles to individually rotate until they are either pointed upwards ($\gamma_k=-\pi/6$) or downwards ($\gamma_k=\pi/6$). In \ding{172}--\ding{174},   the integral curves of the director field outside  triangles oriented at the labelled angles are shown as blue curves, whilst the  effective-boundary defects are shown as red dots,  for $w=100$, $d=5$, $b=1$, $\chi= \pi/4$, and energy-minimizing periods: $\Gamma_1=\Gamma_1^{\min}$, $\Gamma_2=\Gamma_2^{\min}$, and $\Upsilon=\Upsilon^{\min}$.}
    \label{fig:ex2_ContourPlot}
\end{figure}

In the case shown in Fig.~\ref{fig:ex2_sketch}, the three corners that provide a global energy minimum as $w_1,w_2\to\infty$ are the three closest to the horizontal axes passing through the triangle centres (\ie~the preferred alignment of the LC). This heuristic has at least appeared  valid at large separation distances, $d$, and holds asymptotically as $d\to\infty$ (see  \S\ref{sec:farfield_triangle} and App.~\ref{app:ngon}). Thus, the defect locations are predominantly dependent on the orientations of each triangle, \ie~$\chi_1$ and $\chi_2$. Denoting the six corners of the two effective triangles as $\hat{c}_{k}=(1-2/w_1)e^{\im\chi_1}e^{2\im\pi  k/3}$ and $\hat{C}_k= de^{\im\chi}+(b-2/w_2)e^{\im\chi_2}e^{2\im\pi  k/3}$ for  $k\in\{-1,0,1\}$, we find that the three select corners are delineated by the parameter-space plot in  Fig.~\ref{fig:ex2_ContourPlot}.  For example, if $\chi_1=\pi/3$ and $\chi_2=\pi/6$, then the three effective-boundary defects are located at $\hat{c}_{1}$, $\hat{C}_0$, and $\hat{C}_1$, as shown in Fig.~\ref{fig:ex2_sketch} by the red dots. (Note that six out of the eighteen possible corner combinations are never global energy minima.)
For finite anchoring strengths,  the sharp transitions between the twelve regions in Fig.~\ref{fig:ex2_ContourPlot} are smoothed out, with the defects lying close to, but not exactly at the corresponding corners \cite{cs23}. 

\subsection{Body forces and torques}
As in our first example, the dimensionless net force and torque exerted on each of the triangles can be computed  by evaluating the contour integrals in \eqref{eq:complexforcetorque}. Since $\Omega'(z)^2$ is analytic outside the effective triangles, the integration contours can be freely deformed within the liquid crystal bulk. As before, it follows that the force acting on each triangle is equal and opposite, \ie~$(\hatF_2^x, \hatF_2^y)=-(\hatF_1^x, \hatF_1^y)$. We compute the forces and torques integrals using adaptive quadrature  in \textsc{Matlab}.

\begin{figure}[ht!]
    \centering
    \includegraphics[width=0.9\textwidth]{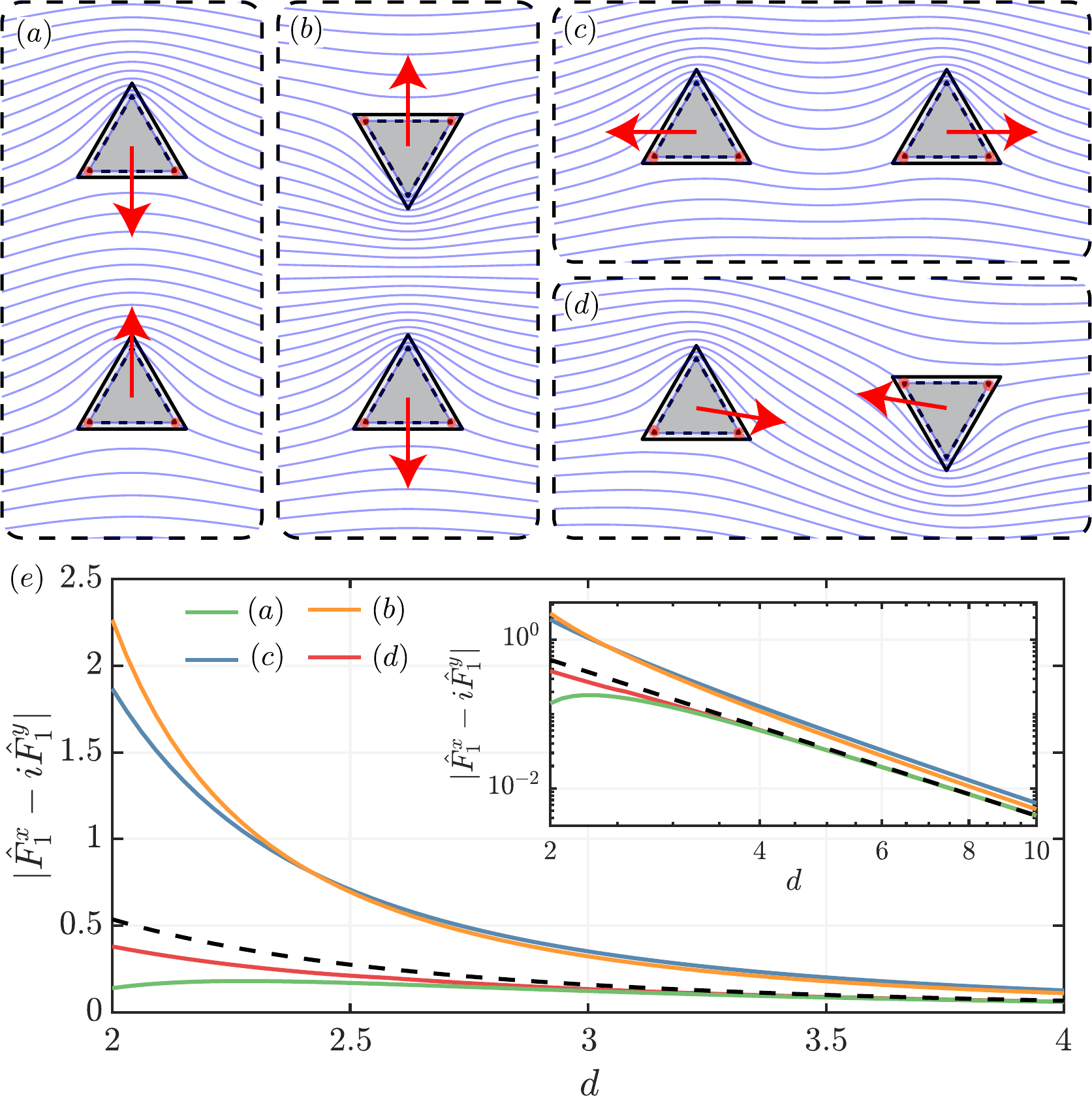}
    \caption{Example 2. Integral curves of the director field outside two identical triangles held in  parallel (a,b)  with $\chi=\pi/2$  and series (c,d) with $\chi=0$, for $d=2$, $b=1$, $w = 10$, and energy minimizing periods: $\Gamma_1=\Gamma_1^{\min}$, $\Gamma_2=\Gamma_2^{\min}$, and $\Upsilon=\Upsilon^{\min}$.  In (a,c) the triangles are pointing in the same direction ($\chi_1=\chi_2=-\pi/6$), whilst in (b,d) the triangles are pointing in opposing directions ($\chi_1=-\chi_2=\pi/6$). The red arrows point in the direction of the force acting on each triangle. The magnitude of the force, $|\hatF_1^x-\im\hatF_1^y|=|\hatF_2^x-\im\hatF_2^y|$, versus the separation distance, $d$, in configurations (a--d)  are plotted as coloured curves in (e). As $d$ increases, all four curves converge to the asymptotic solution \eqref{eq:triangle_asymforce}, which is shown as a black dashed line. The logarithmically-scaled plot is shown in the inset.}
    \label{fig:ex2_UDLR}
\end{figure}

The torques acting on each body desire to rotate the triangles until one of their sides is aligned with the preferred axis of the liquid crystal (\ie~the horizontal, $\chi_k=\pm\pi/6$). Furthermore, the equal and opposite forces acting on each body want  the  triangles to rotate around  a midpoint until they are horizontally ($\chi=0$ or $\pi$) or vertically ($\chi=\pm\pi/2$) aligned. These observations suggest that, if the bodies were free to move and the relaxation time of the liquid crystal was sufficiently small, then the two triangles would ultimately be aligned vertically or horizontally and  pointed upwards or downwards.  Once in this configuration, the interaction force is dependent on the triangles' orientations, as shown in Fig.~\ref{fig:ex2_UDLR}. In particular, we observe  that  in line triangles ($\chi=0$ or $\pi$) \emph{repel} each other when pointed in the same direction (\eg~$\chi_0=\chi_1=-\pi/6$, Fig.~\myref{fig:ex2_UDLR}{a}), but \emph{attract}  when oriented in opposing directions (\eg~$\chi_0=-\chi_1=-\pi/6$, Fig.~\myref{fig:ex2_UDLR}{b}). Inversely,    parallel triangles ($\chi=\pm\pi/2$) \emph{attract} each other when both pointed in the same direction (\eg~$\chi_0=\chi_1=-\pi/6$, Fig.~\myref{fig:ex2_UDLR}{c}), but \emph{repel}  when oriented in opposing directions (\eg~$\chi_0=-\chi_1=\pi/6$, Fig.~\myref{fig:ex2_UDLR}{d}). These numerical results converge to the general far-field asymptotics derived in \S\ref{sec:asymptotics} as $d\to\infty$, in particular \eqref{eq:triangle_asymforce}, as shown in Fig.~\myref{fig:ex2_UDLR}{e}. This orientation-dependent interaction between triangles has previously been observed experimentally by Lapointe et al.~\cite{lms09}.

More will be said about the interactions between two triangular bodies below, but it will first be of use to examine the interactions of two bodies that are well separated more generally.

\section{Far-field interactions between two general bodies}\label{sec:asymptotics}
An isolated body immersed in a liquid crystal only locally disturbs the director field with the director angle decaying according to $\theta\sim \Oh(a/|z|)$ as $|z| \to \infty$ (a dipole) in  general, where $a$ is a length scale associated with the body. If there is no period around  the body, instead the director field decays  more rapidly, as a quadrupole, $\theta\sim \Oh(a^2/|z|^2)$ \cite{cs23}. It follows that two immersed bodies separated by a large distance will only weakly interact. In this section, we shall analyze this weak interaction for two arbitrary  bodies immersed in a director field that is oriented with the $x$-axis in the far-field. The challenge comes from the  periods in \eqref{eq:OmegaPeriod}. Such periods are not possible for an isolated body since it results in a logarithmic growth in the director angle \cite{cs23}. However, this can be avoided in the case of two immersed bodies by introducing opposing periods, as in \eqref{eq:g_period}, including the case in which the second ``body'' is an outer boundary or infinite wall. Importantly, it is thus possible that the solution for an isolated body is not recovered in the large separation distance limit. 

We begin in \S\ref{sec:weak_conformal} by formally introducing the separation distance, $d$, by considering the conformal map, $z=h(\zeta)$, which was introduced in \S\ref{sec:twoimmersedbodies}. Then, in  \S\ref{sec:weak_director}, we derive an expansion for the director field as $d\to\infty$ using the general expression \eqref{eq:director_sol}. Finally, in \S\ref{sec:asymptotic_force} and \S\ref{sec:asymptotic_torque}, we compute the  resulting force and torque acting on each body. These asymptotic results are applied to  the two examples: two immersed cylinders in \S\ref{sec:asymptotic_cylinders} and two immersed triangles in \S\ref{sec:farfield_triangle}

\subsection{Asymptotic conformal map}\label{sec:weak_conformal}
In \S\ref{sec:twoimmersedbodies}, we introduced the conformal map $z=h(\zeta)$, which maps the doubly-connected effective domain, $z\in\partial D_w$, to the annulus, $q\leq |\zeta|\leq 1$, with $\partial D_1^w$ mapped to $|\zeta|=1$, $\partial D_2^w$ mapped to $|\zeta|=q$, and $z=\infty$ mapped to $\zeta=\zeta_\infty\in(q,1)$. By construction, this map is analytic in the annulus except for a first-order pole at $\zeta=\zeta_\infty$.  It can, thus, be expressed as the Laurent expansion
\begin{equation}\label{eq:hexpansion}
    z=h(\zeta) = \frac{C_\infty}{\zeta-\zeta_\infty}+\sum_{j=0}^\infty h_j \zeta^j +\sum_{j=1}^\infty  \frac{H_j}{\zeta^j},
\end{equation}
for some complex coefficients $h_j$ and $H_j$. (For example, for the two cylinders considered in \S\ref{sec:ex1}, $C_\infty = -\rho(w_1) e^{\im\chi}(1-\zeta_\infty^2) $, $h_0=\rho(w_1) e^{\im\chi}\zeta_\infty$, and $h_j=H_j=0$ for $j\geq 1$.)

If we instead consider the two bodies as isolated, then, according to the Riemann mapping theorem,  there exist two conformal maps, $z=a(\ww)$ and $z=A(\ww)$,  from the exterior of $\partial D_1$ and the exterior of $\partial D_2$ to the interior of a unit circle $|\ww|<1$, where $\ww$ is a new complex variable. These maps may always be defined such that $z=\infty$ is mapped to $\ww=0$ and can be written as the series expansions
\begin{equation}\label{eq:aAexpansion}
    z=a(\ww) = \frac{a_{-1}}{\ww}+\sum_{j=0}^\infty  a_j \ww^j\quad\text{and} \quad    z=A(\ww) = \frac{A_{-1}}{\ww}+\sum_{j=0}^\infty  A_j \ww^j,
\end{equation}
for some complex coefficients $a_j$ and $A_j$. In addition, since the disc has rotational symmetry, $a(\ww)$ and $A(\ww)$ both contain a rotational degree of freedom, we shall fix this by assuming $a_{-1}>0$ and $A_{-1}>0$. (For example, for the two cylinders consider in \S\ref{sec:ex1}, $a_{-1}=\rho(w_1)$, $A_{-1}=b\rho(b w_2)$, $a_0=0$, $A_0= de^{\im\chi}$, and $a_j=A_j=0$ for $j\geq 1$.) Here, it useful to introduce the  two parameters
\begin{subequations}
\begin{equation}
    d\coloneqq |A_0-a_0| \quad\text{and}\quad e^{\im\chi} \coloneqq \frac{A_0-a_0}{|A_0-a_0|},\subtag{a,b}
\end{equation}
\end{subequations}
which  measure the bodies' separation distance and relative argument, respectively.

As the distance between the two bodies is increased (\ie~$d\to\infty$), we expect $q/\zeta_\infty\to0$ and $\zeta_\infty\to 0$. Furthermore, the  conformal maps from the doubly-connected domain to the annulus, $h(\zeta)$, and from the doubly-connected domain to the inverted annulus, $h(q/\zeta)$, should  recover the maps $a(\ww)$ and $A(\ww)$ up to rotational degree of freedom. That is,  $h(\zeta)\sim a(e^{\im b}\zeta)$  and $h(q/\zeta)\sim A(e^{\im c}\zeta)$ as $d\to \infty$ for some $b, c\in(-\pi,\pi]$ to be determined. Inserting the series expansions \eqref{eq:hexpansion} and \eqref{eq:aAexpansion} yields the leading-order balance
\begin{subequations}
\begin{align}
 \sum_{j=0}^\infty h_j \zeta^j +\sum_{j=1}^\infty  \frac{H_j+C_\infty \zeta_\infty^{j-1} }{\zeta^j}&\sim \frac{a_{-1}}{e^{\im b}\zeta}+\sum_{j=0}^\infty  a_j \left(e^{\im b  }\zeta\right)^j,\\
  \sum_{j=0}^\infty \frac{(h_j-C_\infty/\zeta_\infty^{j+1}) q^j}{\zeta^j} +\sum_{j=1}^\infty  \frac{H_j \zeta^j}{q^j} &\sim \frac{A_{-1}}{e^{\im c}\zeta}+\sum_{j=0}^\infty  A_j \left(e^{\im c  }\zeta\right)^j,
\end{align}
\end{subequations}
as $d \to \infty$. Above,  we have expanded the pole at $\zeta=\zeta_\infty$ using the fact that $\zeta\ll\zeta_\infty$ and $\zeta\ll q/\zeta_\infty$.

Equating the series coefficients yields the asymptotic expressions for $C_\infty$, $\zeta_\infty$, $q$, $h_j$, and $H_j$ as $d\to \infty$. The unknown rotation angles, $b$ and $c$, are then set by enforcing the requirements that $q>0$ and $\zeta_\infty>0$, which yields $b=\pi-\chi$ and $c=-\chi$. Together, we find that 
\begin{subequations}\label{eq:finalasympotic_conformal}
\begin{gather}
    C_\infty \sim -a_{-1} e^{ \im \chi},\quad\zeta_\infty\sim\frac{a_{-1}}{d} ,\quad q\sim  \frac{a_{-1}A_{-1}}{d^2},\subtag{a--c}\\
         h_j\sim  a_j \left(-e^{-\im \chi}\right)^j,\quad
         H_j \sim A_j \left(q e^{-\im \chi}\right)^j,\subtag{d,e}
\end{gather}
\end{subequations}
as $d\to \infty$. It follows that the conformal map has three dominant behaviours,
\begin{equation}\label{eq:h_cases}
   z= h(\zeta) \sim \begin{cases}
        a(\ww) &\quad\text{for $\ww = -e^{-\im\chi}\zeta$}, \\
        A(\ww)&\quad\text{for $\ww = q e^{-\im\chi}/\zeta$}, \\
        (A_0-a_0)/(1-\ww)+a_0&\quad\text{for $\ww = \zeta/\zeta_\infty$},
    \end{cases}
\end{equation}
as $d\to\infty$,   corresponding to the three regions: (i) local to the primary body, $|\zeta|\sim 1$; (ii) local to the secondary body, $|\zeta|\sim q  =\Oh(1/d^2)$; and (iii) away from both bodies, $|\zeta|\sim \zeta_\infty  =\Oh(1/d)$.

\subsection{Asymptotic director field}\label{sec:weak_director}
In \S\ref{sec:twoimmersedbodies}, the horizontal liquid crystal outside two immersed bodies  was found to  have a complex director angle \eqref{eq:director_sol}. Using the fact that $q=\Oh(1/d^2)$ and $\zeta_\infty=\Oh(1/d)$ as $d\to \infty$, \ie~\eqref{eq:finalasympotic_conformal},   the infinite sum  in \myeqref{eq:PK}{b} can be written as 
\begin{equation}
    K(\zeta)=- \frac{\zeta}{1-\zeta}-\sum_{j=1}^{\infty}\frac{q^{2j}e^{2\im\alpha}\zeta^j}{1-q^{2j}e^{2\im\alpha}}+\sum_{j=1}^{\infty}\frac{q^{2j}e^{-2\im\alpha}/\zeta^j}{1-q^{2j}e^{-2\im\alpha}} ,
\end{equation}
 provided  $1/d^{4}\ll|\zeta|\ll d^{4}$. Inserting this  into \eqref{eq:dF}  yields the potential expansion 
\begin{equation}\label{eq:dF_asym}
\begin{split}
    F'(\zeta)&= \frac{\overline{C_\infty} e^{\im\beta}}{(\zeta_\infty \zeta-1)^2}-\frac{C_\infty e^{-\im\beta}}{(\zeta-\zeta_\infty)^2}  +\frac{1
    }{2\pi\im}\frac{\Gamma_1+\Gamma_2 e^{\im\alpha}}{\zeta-1/\zeta_\infty}-\frac{1
    }{2\pi\im}\frac{\Gamma_1+\Gamma_2  e^{-\im\alpha}}{\zeta-\zeta_\infty}\\
 &\quad+\frac{\Gamma_2 e^{-\im\alpha}}{2\pi\im\zeta}-\sum_{j=1}^{\infty} \frac{q^{2j}}{\zeta_\infty^{j+1}}\left(\overline{F_j} \zeta^{j-1}-\frac{{F_j}}{\zeta^{j+1}}\right),
    \end{split}
\end{equation}
provided $1/d^{3}\ll|\zeta|\ll d^{3}$, with the $\Oh(1)$-constants
\begin{equation}\begin{split}
        F_j &\coloneqq       \frac{ 1}{e^{2\im\alpha}- q^{2j}}\left[j\overline{C_\infty} e^{\im\beta}+ \frac{\Gamma_1+\Gamma_2 e^{\im\alpha}}{2\pi\im}\zeta_\infty\right]\\
        &\quad-
      \frac{ \zeta_\infty^{2j}}{e^{2\im\alpha}- q^{2j}}\left[j C_\infty e^{-\im\beta}+\frac{\Gamma_1+\Gamma_2 e^{-\im\alpha}}{2\pi\im}\zeta_\infty\right].
      \end{split}
\end{equation}

At this point, we can distinguish the three asymptotic regions described in \eqref{eq:h_cases}. In each of these regions, the expression \eqref{eq:dF_asym} can be simplified further and the corresponding  director field, \eqref{eq:director_sol}, can be computed as $d\to\infty$. Below we provide the resulting  complex director field at leading-order:

(i) Local to the primary body, $\ww=-e^{-\im\chi}\zeta=\Oh(1)$, we find that
\begin{equation}\label{eq:asym_primarydirector}
    \Omega(z)= \log\left[\frac{ (1-e^{\im(\beta+\gamma_1)}\ww)(1+e^{\im(\beta-\gamma_1)} \ww)}{-\ww ^2 a'(\ww)/a_{-1}}\right]-\im\beta-\frac{\Upsilon}{2\pi\im}\log \ww+ \Oh(1/d),
\end{equation}
with $\ww =a^{-1}(z)+\Oh(1/d)$ and   $\beta = -\Upsilon\log(d/a_{-1})/(2\pi)+\Oh(1/d)$  as $d\to\infty$, for $\Gamma_1 = 4\pi a_{-1}\sin\gamma_1$.

(ii) Local to the secondary body, $\ww = q e^{-\im\chi}/\zeta=\Oh(1)$,  we find that
\begin{equation}\label{eq:asym_secondarydirector}
    \Omega(z)= \log\left[\frac{ (1-e^{\im(\delta+\gamma_2)}\ww)(1+e^{\im(\delta-\gamma_2)}\ww)}{-\ww^2 A'(\ww)/A_{-1}}\right]-\im\delta+\frac{\Upsilon}{2\pi\im}\log \ww+ \Oh(1/d),
\end{equation}
with $\ww =A^{-1}(z)+\Oh(1/d)$ and $\delta=\beta-\alpha =\Upsilon\log(d/A_{-1})/(2\pi)+\Oh(1/d)$ as $d\to\infty$, for $\Gamma_2 = 4\pi A_{-1}\sin\gamma_2$. 

(iii) Away from both bodies, $\ww=\zeta/\zeta_\infty=\Oh(1)$,  we find that 
\begin{equation}\label{eq:weak_director_farfield}
   \Omega(z)=-\frac{\Upsilon}{2\pi\im}\log \ww+\frac{\Gamma_1 e^{\im\beta} }{2\pi\im}\frac{1-\ww}{d e^{\im\chi}}+\frac{\Gamma_2 e^{\im\delta} }{2\pi\im}\frac{1/\ww-1}{d e^{\im\chi}}+\Oh(1/d^2),
\end{equation}
with $\ww\sim (z-A_0)/(z-a_0)+\Oh(1/d^2)$  as $d\to\infty$.

Overall, as the separation distance increases ($d\to \infty$), the director fields local to the two bodies, \eqref{eq:asym_primarydirector} and \eqref{eq:asym_secondarydirector}, do not recover the director field outside  an isolated body  \cite{cs23}. Instead, they are coupled by an equal and opposite  period $\Upsilon$, which induces a logarithmic director angle away from the two bodies, that is 
\begin{equation}
    \theta(z) =-\Im\Omega(z) \sim \log\left|\frac{z-a_0}{z-A_0}\right| +\Oh(1/d),
\end{equation}
in the far-field, \ie~\eqref{eq:weak_director_farfield}. The period $\Upsilon$, as well as the two other periods $\Gamma_1$ and $\Gamma_2$, also control the locations of the four effective-boundary defects (as introduced in \S\ref{sec:cylinder_defects}), these are located on the effective triangle at $z\sim a(\pm e^{-\im(\beta\pm\gamma_1)})$ and $z\sim A(\pm e^{-\im(\delta\pm\gamma_2)})$. Determining these three periods requires the minimization of the free energy of the liquid crystal.
This energy is given by the boundary integral expression in \eqref{eq:complexenergy}, but takes the form
\begin{equation}\label{eq:asym_energy}
    \hatE= \hatE_1(\Gamma_1,\Upsilon) + \hatE_2(\Gamma_2,\Upsilon)+\Oh(1/d), 
\end{equation}
as $d\to\infty$, where $\hatE_1$ and $\hatE_2$ are the boundary integral in \eqref{eq:complexenergy} around the two bodies $\partial D_1$ and $\partial D_2$ with the director angles \eqref{eq:asym_primarydirector} and \eqref{eq:asym_secondarydirector}, respectively. These integrals can be evaluated and, hence, minimized for  $\Upsilon$,  $\Gamma_1$, and $\Gamma_2$ to $\Oh(1/d)$.

\subsection{Asymptotic  force}\label{sec:asymptotic_force}

The force acting on a body submerged in a liquid crystal is given by \myeqref{eq:complexforcetorque}{a}. Since $\Omega'(z)$ is analytic within the liquid crystal, it follows from  Cauchy's integral theorem that the force acting on the primary body is equal and opposite to the force acting on the secondary body, which can be expressed as 
\begin{equation}
    \hatF^x_1-\im \hatF^y_1=-\hatF^x_2+\im \hatF^y_2= -\frac{1}{2\im} \oint_{|\zeta|=R}\left(\frac{\de\Omega}{\de \zeta}\right)^2
    \frac{\de \zeta }{z'(\zeta)},
\end{equation}
for any $R\in(q,1)$. If we take $R=\zeta_\infty$, then we can insert the asymptotic expansion \eqref{eq:weak_director_farfield} with $\ww=\zeta/\zeta_\infty$,  this yields
\begin{equation} \label{eq:asym_force_integral}
    \hatF^x_1-\im \hatF^y_1\sim \frac{1}{8\pi^2\im de^{\im\chi}} \oint_{|\ww|=1}\left(\frac{\Upsilon}{\ww}+\frac{\Gamma_1 e^{\im\beta}}{d  e^{\im\chi}}+\frac{\Gamma_2e^{\im\delta}}{d  e^{\im\chi}} \frac{1}{\ww^2} \right)^2\left(1-\ww\right)^2\de \ww,
\end{equation}
as $d\to\infty$. The integrand in \eqref{eq:asym_force_integral} is analytic in  $|\ww|\leq 1$ except for a pole at $\ww=0$. Cauchy's residue theorem, thus, yields
\begin{equation}\label{eq:asym_force}
    \hatF^x_1-\im \hatF^y_1=-\hatF^x_2+\im \hatF^y_2\sim -\frac{\Upsilon^2}{2\pi d e^{\im\chi}}+\frac{\Upsilon(\Gamma_1 e^{\im\beta}+\Gamma_2 e^{\im\delta})}{2\pi(d e^{\im\chi})^2} - \frac{ \Gamma_1\Gamma_2e^{\im(\beta+\delta)}}{\pi(d e^{\im\chi})^3},
\end{equation}
as $d\to \infty$, for $\beta =- \Upsilon/(2\pi)\log(d/a_{-1}) $ and $\delta = \Upsilon/(2\pi)\log(d/A_{-1}) $.

\subsection{Asymptotic torque}\label{sec:asymptotic_torque}

The torque acting on a body submerged in the liquid crystal is given by \myeqref{eq:complexforcetorque}{b}. Since $\Omega'(z)$ is analytic within the liquid crystal, it follows from  Cauchy's integral theorem that the torque acting on the primary body is 
\begin{equation}
\hatT_1= \Upsilon -\frac{1}{2}\Re\left[ \oint_{|\zeta|=R}\left(z(\zeta)-a_0\right)\left(\frac{\de\Omega}{\de \zeta}\right)^2\frac{\de \zeta}{z'(\zeta)}\right],
\end{equation}
for any $R\in(q,1)$. If we again take $R=\zeta_\infty$, then we can insert the asymptotic expansion \eqref{eq:weak_director_farfield},  this yields
\begin{equation}
\hatT_1\sim\Upsilon +\frac{1}{8\pi^2}\Re\left[ \oint_{|\ww|=1}\left(\frac{\Upsilon}{\ww}+\frac{\Gamma_1 e^{\im\beta}}{d  e^{\im\chi}}+\frac{\Gamma_2e^{\im\delta}}{d  e^{\im\chi}} \frac{1}{\ww^2} \right)^2\left(1-\ww\right)\de \ww\right].
\end{equation}
The integrand is again analytic in  $|\ww|\leq1$ except for a pole at $\ww=0$. Cauchy's residue theorem, thus, yields
\begin{equation}\label{eq:asym_torque1}
\hatT_1\sim\Upsilon +\frac{\Upsilon\Gamma_1 }{2\pi d}\sin(\chi-\beta)-\frac{\Gamma_1\Gamma_2 }{2\pi d^2}\sin(2\chi-\beta-\delta),
\end{equation}
as $d\to \infty$. The torque on the secondary body is similarly determined to be
\begin{equation}\label{eq:asym_torque2}
\hatT_2\sim -\Upsilon -\frac{\Upsilon\Gamma_2 }{2\pi d}\sin(\chi-\delta)+\frac{\Gamma_1\Gamma_2 }{2\pi d^2}\sin(2\chi-\beta-\delta),
\end{equation}
as $d\to \infty$.

\subsection{Far-field interactions between two cylinders} \label{sec:asymptotic_cylinders}

As a first example of far-field interactions, consider the two cylinders introduced in  \S\ref{sec:ex1}. Here, the asymptotic director field local to $|z|=1$ and $|z-de^{\im\chi}|=b$ is given by \eqref{eq:asym_primarydirector} and \eqref{eq:asym_secondarydirector} with the conformal maps
\begin{subequations}
\begin{equation}
z=a(\ww) = \rho(w_1)/\ww  \quad \text{and} \quad z=A(\ww) =d e^{\im\chi}+ b\rho(bw_2)/\ww ,\subtag{a,b}
\end{equation}
\end{subequations}
respectively.   Due to rotational symmetry of the cylinder,  the energy of the liquid crystal local to each body --- \ie~$\mathcal{E}_1(\Gamma_1,\Upsilon)$ and $\mathcal{E}_2(\Gamma_2,\Upsilon)$ in \eqref{eq:asym_energy} --- is minimized when the four effective-boundary defects,  $z\sim  \pm\rho(w_1)e^{\im(\beta\pm\gamma_1)}$ and  $ z-d e^{\im\chi}\sim\pm b\rho(bw_2)e^{\im(\delta\pm\gamma_2)}$, are  aligned with the preferred axis of the liquid crystal \cite{cs23}. It follows that the  free energy is minimized when $\beta=\delta=\gamma_1=\gamma_2=0$; thus, the leading-order  director fields, \eqref{eq:asym_primarydirector} and \eqref{eq:asym_secondarydirector}, recover the solution for an isolated cylinder \cite{cs23}. Due to the two-fold symmetry of a nematic, the periods must  also be  invariant under the map $de^{\im\chi}\mapsto -d e^{\im\chi}$, so  $\Upsilon=\Oh(1/d^2)$, $\Gamma_1=\Oh(1/d^2)$, and $\Gamma_2=\Oh(1/d^2)$ as $d\to \infty$. Higher-order terms are, thus, needed in \eqref{eq:asym_force} to obtain a leading-order expression for the  force acting on the cylinders.

After accounting for the higher-order terms in \eqref{eq:dF_asym}, we find that 
\begin{equation}
    \hatF^x_1-\im \hatF^y_1=-\hatF^x_2+\im \hatF^y_2\sim-\frac{\Upsilon^2}{2\pi d  e^{\im\chi}}-\frac{2\im\Upsilon(a_{-1}^2-A_{-1}^2)}{  \left(d e^{\im\chi}\right)^3}-\frac{48\pi a_{-1}^2A_{-1}^2}{ \left(d e^{\im\chi}\right)^5},
\end{equation}
as $d\to \infty$, where  $a_{-1}= \rho(w_1)\sim 1-1/w_1$ and $A_{-1}= b\rho(b w_2)\sim b-1/w_2$ are the radii of the two effective cylinders. When the radii of the effective cylinders equate, \ie~$a_{-1}=A_{-1}$,  the added symmetry implies that $\Upsilon=0$ and $\Gamma_1=-\Gamma_2$ for all separation distances, $d>0$. In this  case, the asymptotic force is given by the quadrupolar interaction \eqref{eq:symmetriccylinder_asymforce} with $C=48\pi a_{-1}^2A_{-1}^2$, which is consistent  with the results presented in Fig.~\ref{fig:ex1_Force}.

\subsection{Far-field interactions between two triangles}\label{sec:farfield_triangle}

As a second example, consider the two triangular prisms explored in \S\ref{sec:ex2}. Here, the asymptotic  director fields local to the two triangles $\partial D_1$ and $\partial D_2$ are given by \eqref{eq:asym_primarydirector} and \eqref{eq:asym_secondarydirector}  with the conformal maps
\begin{subequations}\label{eq:farfield_triangle_conformalmap}
\begin{align}
    z =a(\ww) &= \left(1-\frac{2}{w_1}\right)\frac{h( e^{3\im \chi_1} \ww^3)}{h(1)\ww}, \\
\text{and} \quad z =A (\ww) &= d e^{\im\chi}+ \left(b-\frac{2}{w_2}\right)\frac{h( e^{3\im \chi_2} \ww^3)}{h(1)\ww}, 
\end{align}
\end{subequations}
respectively, for the  hypergeometric function $h(\zeta)={}_{2}h_1(-2/3,-1/3;2/3;\zeta)$. The conformal maps $a(\ww)$ and $A(\ww)$ are Schwarz--Christoffel mappings, which map the exterior of the effective polygon onto the unit disc  \cite{ablowitz2003,cs23}. The corners of the effective polygons,  $z=\hat{c}_{k}=(1-2/w_1)e^{\im\chi_1}e^{2\im\pi  k/3}$ and $z=\hat{C}_k= de^{\im\chi}+(b-2/w_2)e^{\im\chi_2}e^{2\im\pi  k/3}$ for $k\in\{-1,0,1\}$, are mapped to  points on the unit circle, $\ww=\hat{b}_k=e^{-\im\chi_1}e^{-2\im\pi  k/3}$ and $\ww=\hat{B}_k=e^{-\im\chi_2}e^{-2\im\pi  k/3}$, respectively.  

For large anchoring strengths ($w_1,w_2\to\infty$), the energy of the liquid crystal attains a local minimum when three of the four effective-boundary defects are located at corners of the effective triangles. Furthermore, the energy is globally minimized when the three  corners are those closest to the horizontal axes passing through the triangle centres, as delineated  in Fig.~\ref{fig:ex2_ContourPlot}.  (This observation is shown to hold asymptotically in App.~\ref{app:ngon}.) Since the locations of the four effective-boundary defects are known asymptotically, \ie~$z\sim a(\pm e^{-\im(\beta\pm\gamma_1)})$ and $z\sim A(\pm e^{-\im(\delta\pm\gamma_2)})$, one can apply a simple geometric argument to derive expressions for the three periods as $d\to\infty$. Below, we present the results of this argument assuming that $|\chi_1|,|\chi_2|\leq\pi/3$, without loss of generality.

If $\left|\left|\chi_1\right|-\pi/6\right|<\left|\left|\chi_2\right|-\pi/6\right|$, then the two effective-boundary defects on $\partial D_1$ lie at corners, whilst the location of the third free defect depends on the orientation of $\partial D_2$. It follows that
\begin{subequations}\label{eq:triangle_asym_period1}
\begin{gather}
     \gamma_1 = \sgn(\chi_1)\pi/6,\qquad \beta = \chi_1-\sgn(\chi_1)\pi/6 , \subtag{a,b}\\
        \gamma_2 = \begin{cases}
        \chi_2-\delta &\quad \text{if $0<|\chi_2|<\pi/6$},\\
        \delta-\chi_2+\sgn(\chi_2)\pi/3 &\quad \text{if $\pi/6<|\chi_2|$}.
    \end{cases}\subtag{c}
\end{gather}
\end{subequations}
Alternatively, if $\left|\left|\chi_1\right|-\pi/6\right|>\left|\left|\chi_2\right|-\pi/6\right|$, then the two effective-boundary defects on $\partial D_2$ lie at corners and we find that
\begin{subequations}\label{eq:triangle_asym_period2}
\begin{gather}
     \gamma_2 = \sgn(\chi_2)\pi/6,\qquad \delta = \chi_2-\sgn(\chi_2)\pi/6 , \subtag{a,b}\\
        \gamma_1 = \begin{cases}
        \chi_1-\beta &\quad \text{if $0<|\chi_1|<\pi/6$},\\
        \beta-\chi_1+\sgn(\chi_1)\pi/3 &\quad \text{if $\pi/6<|\chi_1|$}.
    \end{cases}\subtag{c}
\end{gather}
\end{subequations}
With these variables, the three periods can be computed using $\Gamma_1\sim 4\pi \sin\gamma_1/a_{-1}$, $\Gamma_2\sim 4\pi b \sin\gamma_2/A_{-1}$, and $\Upsilon \sim -2\pi\beta/\log|d/a_{-1}|\sim2\pi\delta/\log|d/A_{-1}|$, for $a_{-1} =(1-2/w_1)/h(1) $ and $A_{-1} =(b-2/w_2)/h(1)$. The asymptotic forces and torques are then given by \eqref{eq:asym_force}, \eqref{eq:asym_torque1}, and \eqref{eq:asym_torque2}. 

As an example,  consider the case of two identical triangles ($b=1$) with large anchoring strengths ($w=w_1=w_2\gg1$). Here, the  torques acting on each triangle satisfy
\begin{equation}\label{eq:triangle_asymtorque}
     \hat{T}_1 \sim  -\hat{T}_2 \sim\frac{2\pi}{\log|d/a_{-1}|}\begin{cases}
    \sgn(\chi_1)\pi/6-\chi_1 \quad \text{if $\left|\left|\chi_1\right|-\pi/6\right|<\left|\left|\chi_2\right|-\pi/6\right|$}, \\
 \chi_2-\sgn(\chi_2)\pi/6 \quad \text{if $\left|\left|\chi_1\right|-\pi/6\right|>\left|\left|\chi_2\right|-\pi/6\right|$},
    \end{cases}
\end{equation}
as $d\to\infty$, where $a_{-1}=(1-2/w)/h(1)\approx 0.73(1-2/w)$. (Note that  this  expression for $\log|d/a_{-1}|\hat{T}_1/(2\pi)$ is shown as the  contours in the $(\chi_1,\chi_2)$-space in Fig.~\ref{fig:ex2_ContourPlot}.) These torques drive the triangles to individually rotate until they either point upwards ($\chi_k=-\pi/6$) or downwards ($\chi_k=\pi/6$), with $\Upsilon=0$. Critically,  the torques in \eqref{eq:triangle_asymtorque} only  decay proportional to  $1/\log d$ as $d\to\infty$, thus one would expect the triangles to experience a rotation even when very well separated. Once  oriented with $\chi_k=\pm \pi/6$, the triangles interact according to the  force
\begin{equation}\label{eq:triangle_asymforce}
    \hatF^x_1-\im \hatF^y_1=-\hatF^x_2+\im \hatF^y_2 = -\sgn(\chi_1\chi_2)\frac{4\pi 
    a_{-1}^2}{(de^{\im\chi})^3} +\Oh(1/d^4),
\end{equation}
as $d\to\infty$, where $a_{-1}=(1-2/w)/h(1)\approx 0.73(1-2/w)$. In particular, this force rotates the triangles around each other until they are in parallel ($\chi=\pm\pi/2$) or series ($\chi=0$ or $\pi$), depending on if they are  pointed in the same direction ($\chi_1=\chi_2=\pm\pi/6$) or in opposite directions ($\chi_1=-\chi_2=\pm\pi/6$), respectively. In either case, the triangles are  then attracted to each other, as observed in Fig.~\myref{fig:ex2_UDLR}{a,d}. It should be noted that the magnitude of the force  in \eqref{eq:triangle_asymforce} is $| \hatF^x_1-\im \hatF^y_1| \sim4\pi a_{-1}^2/d^3$, which is independent of the triangle orientation. This result is compared to the full numerical solutions in Fig.~\myref{fig:ex2_UDLR}{e}, suggesting that one obtains a good approximation provided the bodies are separated by (approximately) two body widths. Shape-dependent interactions have indeed been observed experimentally~\cite{lms09,bgl15}.

\section{Conclusions} \label{sec:conc}

Even though the director angle in a nematic LC is a harmonic function in the single Frank constant approximation, finding solutions is not a simple task. Nonlinear, Robin boundary conditions provide one challenge, but a far greater difficulty lies in the selection of topological defect locations, either on body surfaces in the strong anchoring limit, or on effective boundaries outside of the fluid domain for weak (finite) anchoring strengths. While this was somewhat straight-forward for a single immersed particle \cite{cs23}, multiple bodies demand a more technical analysis. Fortunately, conformal mapping techniques for multiply-connected domains could be used effectively as part of this program \cite{Crowdy20}.

Looking ahead, the equilibrium configuration provides a first step in the direction of modelling the anisotropic viscous drag on moving bodies \cite{rt95,sv01,lhp04,gd13} and the dynamics of bodies immersed in active suspensions \cite{ls22,rzd23,wszs23}, for applications like microrheology \cite{gd16,csmsd16} and self-assembly \cite{wg02} (see also Refs.~\cite{Muvsevivc17,Smalyukh18}). Fluid anisotropy also impacts individual bacterial trajectories \cite{mttwa14,zsla14,tmasw15,ksp15,fdoa19,su23}, as well as the interactions among nearby bacteria \cite{szla15}. The locations of topological defects are of particular interest in an effort to template molecular self-assembly \cite{wmbda16,lsbglyks16}, and their tendency to reside near sharp boundary features is intriguing \cite{bgl15}. The solutions presented herein may offer a degree of insight on these current scientific pursuits. 

Another question of interest, but one which requires different tools to explore, pertains to the relevance of distinct bend and splay moduli. Fortunately, these moduli are comparable in common liquid crystals like PAA, 5CB and DSCG at room temperature \cite{degennes1993,brbf85,znnbsls14}, and we suspect that the changes from the present results will be limited. Twist moduli can be substantially smaller, however, and out-of-plane relaxation of stress is another generic possibility (see for example Ref.~\cite{Williams86}) that should be addressed.

Other open questions are of a more analytical variety. The energy change with body rotations has also been recently considered, and found to be no worse than Lipschitz continuous in the orientation of bodies in three dimensions \cite{ablv23}. With defects jumping from the corners on one body to another under rotation, as we have observed with two triangular bodies in \S\ref{sec:ex2}, we conjecture that no further smoothness in the energy will be possible to show generally. 

Although our examples were restricted to the examination of two bodies, there is nothing in the analysis presented here that does not immediately extend to a greater number of bodies. We are eager to see these techniques used to describe many-body elastic interactions, though it may be that the simpler far-field interactions will prove more useful as a starting point for suspension configurations.

\section*{Acknowledgments} Helpful conversations with Nicholas Abbott, Thomas Powers, and  Raghav Venkatraman are gratefully acknowledged. SES acknowledges the UW--Madison Office of the Vice Chancellor for Research and Graduate Education and funding from the Wisconsin Alumni Research Foundation.

\appendix

\section{Analytical potentials for a doubly-connected domain}\label{app:potential} 
In this section, we derive analytical solutions to the potential problems \eqref{eq:problem_G} and \eqref{eq:problem_F} by introducing two functions, $G_1(\zeta)$ and $G_2(\zeta)$, that are holomorphic in $q\leq\zeta\leq 1$ except for a logarithmic singularity at $\zeta=a$. The branch cut of this logarithmic singularity is chosen to give a unit period  across $|\zeta|=1$, for $G_1(\zeta)$, or $|\zeta|=q$, for $G_2(\zeta)$. The imaginary part of $G_1$ and $G_2$ are analogous to Green's functions since they are  harmonic except for a logarithmic singularity at $\zeta=a$, they have thus been coined `modified Green's functions' \cite{Crowdy20}. We begin by deriving analytic formulae for the two modified Green's functions using the method of images. We shall then express the wanted potentials, $F$ and $G$, in terms of $G_1$ and $G_2$.

\subsection{First modified Green's function}\label{sec:zerothGreen}
The first modified Green's function, $G_1(\zeta;a,\alpha)$, is defined to be the solution to
\begin{subequations}\label{eq:ZerothGreenProblem}
\begin{align}
G_1(\zeta;a,\alpha) \text{ locally holomorphic} &\qquad \text{ in $q< \zeta< 1$,}\\
 \Im G_1(\zeta)=0 &\qquad \text{ on $|\zeta|=1$,}\\
  \Im\left[e^{\im\alpha} G_1(\zeta)\right]=C_1&\qquad \text{ on $|\zeta|=q$,}\\
G_1(\zeta)\sim \frac{1}{2\pi\im}\log(\zeta-a)&\qquad \text{ as $\zeta\to a$,}\\
\oint_{|\zeta|=1} \de G_1 = 1 \quad & \text{and} \quad \oint_{|\zeta|=q}e^{\im\alpha} \de G_1 = 0, \subtag{e,f}
\end{align}
\end{subequations}
for some unknown constant $C_1$, a given  complex constant $a$, and  given real constants $\alpha$ and $q$.

The  periods in \myeqref{eq:ZerothGreenProblem}{e,f} imply that the branch cut of the logarithmic singularity at $\zeta=a$, \ie~\myeqref{eq:ZerothGreenProblem}{d},  must cross $|\zeta|=1$. To achieve this, we shall first introduce an image of $\log(\zeta-a)/(2\pi\im)$ across $|\zeta|=1$. Using the Schwarz function $\overline{\zeta} = 1/\zeta$ with \myeqref{eq:ZerothGreenProblem}{b} yields $G_1(\zeta) = \overline{G}_1(1/\zeta)$ on $|\zeta|=1$, thus $G_1(\zeta)\sim -\log(\zeta-1/\bar{a})/(2\pi\im)$ as $\zeta\to 1/\bar{a}$. (This is an example of Schwarz reflection principle \cite{ablowitz2003}.) Adding this to the singularity at $\zeta=a$ yields 
\begin{equation}
G_1(\zeta;a,\alpha) = \frac{1}{2\pi\im}\log \frac{\zeta-a}{\zeta-1/\bar{a}} + \text{analytic function,}
\end{equation}
in $q\leq \zeta\leq 1/q$, which  has branch cut  between $\zeta=a$ and $\zeta =1/\bar{a}$. Although this expression satisfies the boundary condition on $|\zeta|=1$, \ie~\myeqref{eq:ZerothGreenProblem}{b}, up to an additive constant, it does not satisfy the boundary condition on $|\zeta|=q$, \ie~\myeqref{eq:ZerothGreenProblem}{c}. To fix this, we shall now introduce its image  across $|\zeta|=q$.

Using the Schwarz function $\overline{\zeta} = q^2/\zeta$ with \myeqref{eq:ZerothGreenProblem}{c}  yields $ G_1(\zeta)= e^{-2\im\alpha}\overline{G}_1(q^2/\zeta)+ \mathrm{const.}$ on $|\zeta|=q$. By a similar argument to the above, it follows that
\begin{equation}
G_1(\zeta;a,\alpha) = \frac{1}{2\pi\im}\log \frac{\zeta-a}{\zeta-1/\bar{a}} +\frac{e^{-2\im\alpha}}{2\pi\im}\log \frac{\zeta-q^2 a}{\zeta-q^2/\bar{a}} + \text{analytic function,}
\end{equation}
in $q^3\leq |\zeta|\leq 1/q$, which has  branch cuts between $\zeta=a$ and $\zeta=1/\bar{a}$ and   $\zeta=q^2 a$ and $\zeta=q^2/\bar{a}$. This expression now satisfies the boundary condition on $|\zeta|=q$, \ie~\myeqref{eq:ZerothGreenProblem}{c}, but the newly added term does not satisfy the boundary condition on $|\zeta|=1$, \ie~\myeqref{eq:ZerothGreenProblem}{b}, thus we now  introduce  its image  across $|\zeta|=1$.

To reflect across $|\zeta|=1$, we again use the fact that $G_1(\zeta) = \overline{G}_1(1/\zeta)$ on $|\zeta|=1$. It follows that
\begin{equation}
\begin{split}
G_1(\zeta;a,\alpha) &= \frac{1}{2\pi\im}\log \frac{\zeta-a}{\zeta-1/\bar{a}} +\frac{e^{-2\im\alpha}}{2\pi\im}\log \frac{\zeta-q^2 a}{\zeta-q^2/\bar{a}}\\
&\quad~ + \frac{e^{2\im\alpha}}{2\pi\im}\log \frac{\zeta-q^{-2} a}{\zeta-q^{-2}/\bar{a}}+\text{analytic function,}
\end{split}
\end{equation}
in $q^3\leq |\zeta|\leq 1/q^3$, which has  branch cuts between $\zeta=a$ and $\zeta=1/\bar{a}$, $\zeta=q^2 a$ and $\zeta=q^2/\bar{a}$, and $\zeta=q^{-2} a$ and $\zeta=q^{-2}/\bar{a}$.

Repeating this argument yields an infinite series of logarithmic cuts:
\begin{equation}
\begin{split}
G_1(\zeta;a,\alpha) &= \frac{1}{2\pi\im}\sum_{k=1}^\infty\left(e^{2k\im\alpha}\log \frac{1/\zeta-q^{2k}/a}{1/\zeta-q^{2k}\bar{a}}+e^{-2k\im\alpha}\log \frac{\zeta-q^{2k} a}{\zeta-q^{2k}/\bar{a}}\right)\\
& \quad~+\frac{1}{2\pi\im}\log\left(\frac{1}{|a|} \frac{\zeta-a}{\zeta-1/\bar{a}}\right).
\end{split}
\end{equation}
Here, we have fixed the additive constant to ensure that $\Im G_1(\zeta)=0$ on $|\zeta|=1$ and the series converges. Note that this  final expression has  branch cuts between $\zeta=q^{2k}a$ and $\zeta = q^{2k}/\bar{a}$ for all $k\in \mathbb{Z}$. It can also be written in the more compact form 
\begin{equation}\label{eq:ZerothGreens}
G_1(\zeta;a,\alpha) = \frac{1}{2\pi\im}\log\left(|a|\frac{P(\zeta/a;\alpha)}{P(\bar{a}\zeta;\alpha)}\right),
\end{equation}
where
\begin{equation}\label{eq:Pfunc}
P(\zeta;\alpha)\coloneqq (1-\zeta)\prod_{k=1}^\infty (1-q^{2k}\zeta)^{e^{2k\im \alpha}}(1-q^{2k}/\zeta)^{e^{-2k\im \alpha}}.
\end{equation}

\subsection{Second modified Green's function}
The second modified Green function, $G_2(\zeta;a,\alpha)$, is defined to be the solution to
\begin{subequations}\label{eq:FirstGreenProblem}
\begin{align}
G_2(\zeta;a,\alpha) \text{ locally holomorphic} &\qquad \text{ in $q< \zeta< 1$,}\\
 \Im G_2(\zeta)=0 &\qquad \text{ on $|\zeta|=1$,}\\
  \Im\left[e^{\im\alpha} G_2(\zeta)\right]=C_2&\qquad \text{ on $|\zeta|=q$,}\\
G_2(\zeta)\sim \frac{e^{-\im\alpha}}{2\pi\im}\log(\zeta-a)&\qquad \text{ as $\zeta\to a$,}\\
\oint_{|\zeta|=1} \de G_2 = 0 \quad & \text{and} \quad \oint_{|\zeta|=q}e^{\im\alpha} \de G_2 = -1, \subtag{e,f}
\end{align}
\end{subequations}
for some unknown constant $C_2$, given complex constant $a$, and given real constants $\alpha$ and $q$. To construct $G_2$, we apply the same  method of images argument used to construct $G_1$ in \S\ref{sec:zerothGreen}.

First, the choice of periods in \myeqref{eq:FirstGreenProblem}{e,f} implies  the branch cut of the logarithmic singularity at $\zeta=a$, \ie~\myeqref{eq:FirstGreenProblem}{d}, must cross $|\zeta|=q$. We, thus,  introduce an image across $|\zeta|=q$ using the fact that $G_2(\zeta) = \overline{G}_2(q^2/\zeta)+\mathrm{const.}$ on $|\zeta|=q$. This yields
\begin{equation}
G_2(\zeta;a,\alpha) = \frac{e^{-\im\alpha}}{2\pi\im}\log \frac{\zeta-a}{\zeta-q^2/\bar{a}} + \text{analytic function,}
\end{equation}
in $q^2\leq \zeta\leq 1$, which has  a  branch cut between $\zeta=a$ and $\zeta=q^2/\bar{a}$. 

Next, we reflect the solution across $|\zeta|=1$ using the fact that $ G_2(\zeta)= \overline{G}_2(1/\zeta)$ on $|\zeta|=1$. This yields
\begin{equation}
G_2(\zeta;a,\alpha) = \frac{e^{-\im\alpha}}{2\pi\im}\log \frac{\zeta-a}{\zeta-q^2/\bar{a}} +\frac{e^{\im\alpha}}{2\pi\im}\log \frac{\zeta-q^{-2}a}{\zeta-1/\bar{a}} + \text{analytic function,}
\end{equation}
in $q^2\leq |\zeta|\leq 1/q^2$, which has  branch cuts between $\zeta=a$ and $\zeta=q^2/\bar{a}$ and $\zeta= q^{-2}a$ and $\zeta=1/\bar{a}$.

Repeating this reflection argument yields
\begin{equation}\label{eq:FirstGreens}
\begin{split}
G_2(\zeta;a,\alpha) &= \frac{e^{-\im\alpha}}{2\pi\im}\sum_{k=1}^\infty\left(e^{2k\im\alpha}\log \frac{1/\zeta-q^{2k}/a}{1/\zeta-q^{2k-2}\bar{a}}+e^{-2k\im\alpha}\log \frac{\zeta-q^{2k} a}{\zeta-q^{2k+2}/\bar{a}}\right)\\
& \quad~+\frac{e^{-\im\alpha}}{2\pi\im}\log\left( \frac{\zeta-a}{\zeta-q^2/\bar{a}}\right),
\end{split}
\end{equation}
where the additive constant is fixed to ensure convergence and $\Im \left[ G_2(\zeta)\right]=0$ on $|\zeta|=1$. Note that this solution has  branch cuts between $\zeta=q^{2k}a$ and $\zeta = q^{2k+2}/\bar{a}$ for all $k\in \mathbb{Z}$. It can also be written in the more compact form 
\begin{equation}
G_2(\zeta;a,\alpha) =\frac{e^{-\im\alpha}}{2\pi\im}\log\left(\frac{|a|^2}{q^2}\frac{P(\zeta/a;\alpha)}{P(\bar{a}\zeta/q^2;\alpha)}\right),
\end{equation}
where $P(\zeta;\alpha)$ is given by \eqref{eq:Pfunc}. 

A special  case worth mentioning is when  $\alpha=0$. For such an $\alpha$, the modified Green's functions, $G_1$ and $G_2$, correspond to those derived by \eg~Crowdy \cite{Crowdy20}. Furthermore,  since $P(\zeta;\alpha)= -\zeta P(q^2\zeta;\alpha)^{e^{2\im\alpha}}$, we find  that
\begin{equation}\label{eq:alpha1iden}
G_1(\zeta;a,0)-G_2(\zeta;a,0) = \frac{1}{2\pi\im}\log\zeta,
\end{equation}
up to an additive real constant. This identity does not hold for $\alpha\neq 0$, however.

\subsection{Derivatives of the modified Green's functions}
By construction, the modified Green's functions, $G_1(\zeta;a,\alpha)$ and $G_2(\zeta;a,\alpha)$, have logarithmic singularities at $\zeta=a$. To obtain  other singularity types, one can take  derivatives with respect to $a_x=\Re a$ or $a_y=\Im a$. For example, here we shall create  a  solution with a first-order pole at $\zeta=a$ by considering the first derivatives.

Consider the following linear combination of  derivatives 
\begin{equation}\label{eq:derivativeGreen}
    \hat{G}(\zeta;a,\alpha,b) = 2\pi\left( b_y\frac{\partial G_1}{\partial a_x}-b_x\frac{\partial G_1}{\partial a_y}\right),
\end{equation}
for some complex constant $b=b_x+\im b_y$. It follows from  \eqref{eq:ZerothGreenProblem} that  $\hat{G}(\zeta;a,\alpha, b)$ satisfies  
\begin{subequations}\label{eq:derGreenProblem}
\begin{align}
\hat{G}(\zeta) \text{ locally holomorphic} &\qquad \text{ in $q< \zeta< 1$,}\\
 \Im \hat{G}(\zeta)=0 &\qquad \text{ on $|\zeta|=1$,}\\
  \Im\left[e^{\im\alpha} \hat{G}(\zeta)\right]=\hat{C}&\qquad \text{ on $|\zeta|=q$,}\\
\hat{G}(\zeta)\sim \frac{b}{\zeta-a}&\qquad \text{ as $\zeta\to a$,}\\
\oint_{|\zeta|=1} \de\hat{G} = 0 \quad & \text{and} \quad \oint_{|\zeta|=q}e^{\im\alpha} \de\hat{G} = 0, \subtag{e,f}
\end{align}
\end{subequations} 
where $\hat{C}\coloneqq 2\pi( c_x\de C_1/\de a_y-  c_y \de C_1/\de a_x)$. In particular, $\hat{G}$ has a first-order pole at $\zeta=a$, \ie~\myeqref{eq:derGreenProblem}{d}.

Using the Wirtinger derivatives \cite{ablowitz2003},
\begin{subequations}
\begin{equation}
    \frac{\partial}{\partial a} =\frac{1}{2}\left(\frac{\partial}{\partial a_x} -\im\frac{\partial}{\partial a_y}  \right) \quad \text{and} \quad    \frac{\partial}{\partial \bar{a}} =\frac{1}{2}\left(\frac{\partial}{\partial a_x} +\im\frac{\partial}{\partial a_y} \right),\subtag{a,b}
\end{equation}
\end{subequations}
\eqref{eq:derivativeGreen} can be written as
\begin{equation}
    \hat{G}(\zeta;a,\alpha,b) =2\pi\im \left( \bar{b} \frac{\partial G_1}{\partial \bar{a}}-b\frac{\partial G_1}{\partial a}\right).
\end{equation}
Inserting the  formula for $G_1(\zeta;a,\alpha)$ derived earlier, \ie~\eqref{eq:ZerothGreens}, we then find that
\begin{equation}\label{eq:DerivGreens}
\hat{G}(\zeta;a,\alpha,b) =\frac{b}{a}K(\zeta/a;\alpha)- \frac{\bar{b}}{\bar{a}}K(\bar{a}\zeta;\alpha)- \frac{b}{2a}+\frac{\bar b}{2\bar a},
\end{equation}
for the infinite sum
\begin{equation}\label{eq:Kfunc}
    K(\zeta;\alpha)\coloneqq \frac{\zeta P'(\zeta;\alpha)}{P(\zeta;\alpha)}  \equiv-\frac{\zeta}{1-\zeta} +\sum_{k=1}^{\infty} \left(\frac{e^{-2k\im\alpha} q^{2k}}{\zeta-q^{2k}}-\frac{e^{2k\im\alpha}  q^{2k}}{1/\zeta-q^{2k}}\right).
\end{equation}

\subsection{Constructing the potentials}
Equipped with the two modified Green's functions, $G_1(\zeta;a,\alpha)$ and $G_2(\zeta;a,\alpha)$, and the first derivative  function, $\hat{G}(\zeta;a,\alpha,b)$, we are  able to construct the two potentials, $G(\zeta)$ and $F(\zeta)$. To do this, we shall use the linearity of the systems \eqref{eq:ZerothGreenProblem}, \eqref{eq:FirstGreenProblem}, and \eqref{eq:derGreenProblem} to construct solutions to the  problems   \eqref{eq:problem_G} and \eqref{eq:problem_F}.

The first potential is given by
\begin{equation}
    G(\zeta) = -\Upsilon \big[G_1(\zeta;\zeta_\infty,1)-G_2(\zeta;\zeta_\infty,1)\big]= - \frac{\Upsilon}{2\pi\im}\log \zeta,
\end{equation}
where we have used the identity \eqref{eq:alpha1iden}. We, hence, find the unknown constants in \eqref{eq:problem_G} to be
\begin{subequations}\label{eq:app_alphabeta}
\begin{equation}
   \alpha= \frac{\Upsilon}{2\pi}\log q \quad \text{and} \quad \beta =  \frac{\Upsilon}{2\pi}\log \zeta_\infty.\subtag{a,b}
\end{equation}
\end{subequations}

The second potential is given by
\begin{equation}
    F(\zeta)= \hat{G}(\zeta;\zeta_\infty,\alpha, C_\infty e^{-\im\beta}) -\Gamma_1 G_1(\zeta;\zeta_\infty, \alpha) -\Gamma_2 G_2(\zeta;\zeta_\infty, \alpha),
\end{equation}
for the $\alpha$ and $\beta$ in \eqref{eq:app_alphabeta}. Inserting \eqref{eq:ZerothGreens}, \eqref{eq:FirstGreens}, and \eqref{eq:DerivGreens} yields
\begin{equation}
\begin{split}
    F(\zeta)&= \frac{C_\infty e^{-\im\beta}}{\zeta_\infty}K(\zeta/\zeta_\infty) -\frac{\overline{C_\infty}e^{\im\beta}}{\zeta_\infty}K(\zeta_\infty\zeta)\\
    &\quad-\frac{\Gamma_1
    }{2\pi\im}\log\frac{P(\zeta/\zeta_\infty)}{P(\zeta_\infty\zeta)} - \frac{\Gamma_2 e^{-\im\alpha} }{2\pi\im}\log\frac{P(\zeta/\zeta_\infty)}{P(\zeta_\infty\zeta/q^2)},
    \end{split}
\end{equation}
to an additive constant. Here, $P(\zeta)$ and $K(\zeta)$ are defined in \eqref{eq:Pfunc} and \eqref{eq:Kfunc}, respectively. It is worth noting that the identity $P(\zeta/q^2)= -\zeta P(\zeta)^{e^{2\im\alpha}}/q^2$ also yields the alternative form
\begin{equation}
 \begin{split}
     F(\zeta)&= \frac{C_\infty e^{-\im\beta}}{\zeta_\infty}K(\zeta/\zeta_\infty) -\frac{\overline{C_\infty}e^{\im\beta}}{\zeta_\infty}K(\zeta_\infty\zeta)+ \frac{\Gamma_2 e^{-\im\alpha} }{2\pi\im}\log\zeta\\
     &\quad-\frac{\Gamma_1+\Gamma_2 e^{-\im\alpha}
     }{2\pi\im}\log P(\zeta/\zeta_\infty) +\frac{\Gamma_1+\Gamma_2 e^{\im\alpha} }{2\pi\im}\log P(\zeta_\infty\zeta).
    \end{split}
 \end{equation}

\section{Energy of the far-field interactions between two triangles}\label{app:ngon}
In this section, we shall evaluate and   minimize the free energy associated with the far-field interactions between two triangles, $\partial D_1$ and $\partial D_2$, immersed in a liquid crystal, as discussed in \S\ref{sec:ex2} and \S\ref{sec:farfield_triangle}. For clarity, we shall concentrate on the case when the two  triangles are identical with $b=1$ and $w=w_1=w_2\gg1$, however we expect the results   to generalize based on numerical observation. Here, we shall  denote the corners of the physical triangles, $\partial D_1$ and $\partial D_2$, as $z=c_{k}=e^{\im\chi_1}e^{2\im\pi  k/3}$ and $z=C_k= de^{\im\chi}+e^{\im\chi_2}e^{2\im\pi  k/3}$ for $k\in\{-1,0,1\}$, respectively. Additionally, the corners of the effective triangles, $\partial D_1^w$ and $\partial D_2^w$, are denoted $z=\hat{c}_{k}=(1-2/w)e^{\im\chi_1}e^{2\im\pi  k/3}$ and $z=\hat{C}_k= de^{\im\chi}+(1-2/w)e^{\im\chi_2}e^{2\im\pi  k/3}$ for $k\in\{-1,0,1\}$, respectively.

As the distance between the triangles increases ($d\to \infty$), the complex director angles local to $\partial D_1$ and $\partial D_2$ are given by \eqref{eq:asym_primarydirector} and \eqref{eq:asym_secondarydirector}, with the conformal maps \eqref{eq:farfield_triangle_conformalmap}. These conformal mappings map the corners of the effective triangles, $z=\hat c_k$ and $z=\hat C_k$,  to points on the unit circle, $\ww=\hat b_k = e^{-\im \chi_1}e^{-2\pi\im k/3}$ and $\ww=\hat B_k = e^{-\im \chi_2}e^{-2\pi\im k/3}$, respectively. Additionally, the corners of the physical triangles, $z= c_k$ and $z= C_k$, are  mapped to  points in the  unit disc,  $\ww= b_k = a^{-1}(c_k)$ and $\ww= B_k = A^{-1}(C_k)$, respectively.

The energy of the coupled system is given by \eqref{eq:asym_energy} as $d\to\infty$, with the contour integral
\begin{equation}\label{eq:energy_poly}
    \hat{\mathcal{E}}_\ind = \frac{1}{4}\int_{\partial D_\ind}\Im\left[ \left(\Omega_\ind(z)- \overline{\Omega_\ind( z)}\right) \Omega_\ind'(z)z_s\right]+ w\Re\left[1-\e^{\Omega_\ind(z)-\overline{\Omega_\ind( z)}}z_s^2\right]\de s,
\end{equation}
where $\Omega_\ind(z)$ is the complex director angle local to $\partial D_\ind$, \ie~\eqref{eq:asym_primarydirector} for $k=1$ and \eqref{eq:asym_secondarydirector} for $k=2$. For large anchoring strengths ($w\to\infty$),  the Rapini--Papoular surface energy, \ie~the term proportional to $w$ in  \eqref{eq:energy_poly}, is $\Oh(1/w)$ --- this can most easily  be seen by combining \eqref{eq:weak_anchor} and \eqref{eq:totalenergy_dimensionless} in the main text. Thus, the  boundary integral  \eqref{eq:energy_poly} takes the simplified form
\begin{equation}\label{eq:energy_poly2}
    \hatE_k(\Gamma_k,\Upsilon) = \frac{1}{2}\int_{\partial D_k}\Im\left[ \Omega_k(z)\right]\Re\left[ \Omega_k(z)\right]_s\de s +\Oh(1/w),
\end{equation}
as $w\to\infty$. The challenge is  to  evaluate this simplified boundary integral as $w\to \infty$. We begin by considering the triangle centred at $z=0$.

\subsection{Primary triangle}
The complex director angle local  to $\partial D_1$ can be written as \eqref{eq:asym_primarydirector}, that is
\begin{equation}\label{eq:Omega_poly1}
    \Omega_1(z)= \log\left[\frac{ (1-e^{\im(\beta+\gamma_1)}\ww)(1+e^{\im(\beta-\gamma_1)} \ww)}{(1-e^{3\im  \chi_1} \ww^3)^{2/3} }\right]-\im\beta\left(1+\frac{\log\ww}{\log(d/a_{-1})}\right),
\end{equation}
with $\ww= a^{-1}(z)$, $\beta =- \Upsilon\log(d/a_{-1})/(2\pi)$, $\Gamma_1=4\pi a_{-1}\sin\gamma_1$, and $a_{-1}=(1-2/w)/h(1)$. This director angle is  singular at  the corners of the effective triangle, $\ww = \hat{b}_k$, and at two effective-boundary defects $\ww=\hat b_\pm= \pm e^{-\im(\beta\pm\gamma_1)}$, which are located at $z=\hat c_\pm=a(\hat b_\pm)$ on the effective triangle. As $w\to\infty$, the effective-boundary defects tend to points on the physical triangle, we shall denote these  points as $z=c_\pm=a(b_\pm)$ so that $\hat{c}_\pm \to c_\pm $ and $\hat{b}_\pm \to b_\pm $ as $w\to\infty$.

In the strong anchoring limit ($w\to\infty$), the physical and effective boundaries are identical and the director field lies tangent to the polygon, $\partial D_1$. It follows that the director angle, $\theta=-\Im\Omega_1$, is  piecewise-constant on the triangle with anticlockwise jumps of $-\pi$ across the effective-boundary defects, $z=\hat{c}_\pm$, and $2\pi/3$ across  the corner defect, $z=\hat c_\ind$.  Inserting \eqref{eq:Omega_poly1} into \eqref{eq:energy_poly2}, thus, yields
\begin{equation}\label{eq:asymptoticE1}
      \hat{\mathcal{E}}_1(\Gamma_1,\Upsilon) \sim\frac{\pi}{3}\left[T(b_{-1})+T(b_0)+ T(b_1)\right]-\frac{\pi}{2}
      \left[T(b_+)+T(b_-)\right]+\frac{\pi \beta^2}{\log\left|h(1)d\right|},
\end{equation}
as $w\to\infty$, for the real-valued function
\begin{equation}
    T(\ww) \coloneqq \log\left|\frac{(1-e^{\im(\beta+\gamma_1)}\ww)(1+e^{\im(\beta-\gamma_1)}\ww)}{\left(1- e^{3\im \chi_1}\ww^{3}\right)^{2/3}}\right|,
\end{equation} 
and periods $\Upsilon\sim -\beta/\log[h(1)d]$ and  $\Gamma_1\sim 4\pi\sin\gamma_1/h(1)$. The last step is to evaluate   $T(b_\pm)$ and $T(b_\ind)$  as $w\to \infty$. The difficulty comes from the fact that these values are singular in this limit since $b_\ind\to \hat{b}_\ind= e^{-\im  \chi_1}e^{-2\pi\im\ind/3}$  and  $b_\pm\to \hat{b}_\pm =\pm e^{-\im(\beta\pm\gamma_1)}$ as $w\to \infty$. We, thus, require the first-order corrections to $b_\ind$ and $b_\pm$ as $w\to\infty$, which we find now.

\paragraph{Corner defects, $z=c_\ind$} An asymptotic expansion of the  conformal map, \myeqref{eq:farfield_triangle_conformalmap}{a}, local to the corners yields 
\begin{equation}
    \frac{c_\ind}{\hat c_\ind} = \frac{a(b_\ind)}{a(\hat b_\ind)}\sim 1 +\frac{3^{5/3}}{5 h(1)}\left(1-b_\ind/\hat b_\ind\right)^{5/3},
\end{equation}
as $b_\ind\to \hat b_\ind$ (\ie~as $w\to\infty$).  Inserting   $ c_\ind/\hat c_\ind= 1/(1-2/w) $   and balancing the first-order terms yields the correction to $b_\ind$ as $w\to \infty$:
\begin{equation}\label{eq:cornercorrection}
\frac{b_\ind}{\hat b_\ind}\sim 1-\frac{10^{3/5}h(1)^{3/5}}{3 w^{3/5}}.
\end{equation}

\paragraph{Effective-boundary defects, $z=c_\pm$} If $z=c_\pm$ does not lie at a corner,  there exists  $\phi_\pm\in(0, \pi)$ and $k_\pm \in\{-1,0,1\}$ such that $\hat{b}_\pm = e^{-\im\chi_1}e^{-2\im (\phi_\pm+\pi k_\pm)/3}$. It then follows from an asymptotic expansion of the conformal map, \myeqref{eq:farfield_triangle_conformalmap}{a}, that 
\begin{equation}
    c_\pm = a(b_\pm) \sim a(\hat{b}_\pm)+e^{\im\chi_1}e^{\im (1+2k_\pm)\pi/3}\frac{(2\sin\phi_\pm )^{2/3}}{h(1)} (1-b_\pm/\hat{b}_\pm),
\end{equation}
as $b^\pm\to \hat b^\pm$ (\ie~as $w\to\infty$).  Inserting $c_\pm\sim\hat{c}_\pm+ e^{\im\chi_1}e^{\im (1+2k_\pm)\pi/3}/w$  and balancing the dominant terms yields  the  correction to $b_\pm$ as $\w\to\infty$:
\begin{equation}
 \frac{b_\pm}{\hat{b}_\pm}\sim 1- \frac{h(1)/w}{(2\sin \phi_\pm)^{2/3}}.
\end{equation}

With these corrections, we are  able to compute  $T(b_k)$ and  $T(b_\pm)$ asymptotically as $w\to\infty$. We find that there are four possible cases, depending on the positions of the effective-boundary defects:

(Case 1: Neither effective-boundary defect lies at a corner.) Here, $\phi_\pm\in(0,\pi)$, which yields
\begin{subequations}
\begin{align}
    T(b_k) &\sim \log\left|\frac{4\sin\left[\left(\phi_++\pi k_+-\pi k\right)/3\right]\sin\left[\left(\phi_-+\pi k_--\pi k\right)/3\right]}{\left[10h(1)/w\right]^{2/5}}\right|,\\
    T(b_\pm) &\sim \log\left|\frac{2h(1) \sin\left[\left(\phi_+-\phi_-+\pi k_+-\pi k_-\right)/3\right]}{w(2\sin \phi_\pm) ^{4/3}}\right|,
\end{align} 
\end{subequations}
as $w\to\infty$.

(Case 2: One effective-boundary defect lies at a corner.) Here, $\phi_+ = 0$ and $\phi_-\in(0,\pi)$. The expression for $T(b_{k_+})$ and $T(b_+)$ are  instead given by 
\begin{equation}
    T(b_{k_+})= T(b_+) \sim \log\left|\frac{2}{3}\sin\left[\left(\phi_- +\pi k_--\pi k_+\right)/3\right]\right|-\frac{1}{5}\log\left|\frac{w}{10h(1)}\right|,
\end{equation} 
as $w\to\infty$. 

(Case 3: Both effective-boundary defects lie at the same corner.) Here, $\phi_+ = \phi_- = 0$ and $k_+= k_-$. The expressions for $T(b_{k_\pm})$ and $T(b_\pm)$ are instead  given by 
\begin{equation}
    T(b_{k_{\pm}})= T(b_\pm) \sim -\frac{4}{5}\log\left|\frac{w}{10h(1)}\right|-2\log 3,
\end{equation} 
as $w\to\infty$. 

(Case 4: Both effective-boundary defects lie at distinct corners.) Here, $\phi_+ = \phi_- = 0$ and $k_+\neq k_-$. The expressions for $T(b_{k_\pm})$ and $T(b_\pm)$ are instead given by 
\begin{equation}
    T(b_{k_{\pm}})= T(b_\pm) \sim -\frac{1}{5}\log\left|\frac{w}{10h(1)}\right|-\frac{1}{2}\log 3,
\end{equation} 
as $w\to\infty$. 

In each of these cases, the energy of the director field local to primary body, \ie~\eqref{eq:asymptoticE1}, takes the form
\begin{equation}\label{eq:E1_asymp}
    \hat{\mathcal{E}}_1(\Gamma_1,\Upsilon)\sim \pi A_1\log\left|\frac{ w }{10h(1)}\right|+\pi B_1+\frac{\pi \beta^2}{\log\left|h(1)d\right|},
\end{equation}
as $w\to \infty$, with the coefficients
\begin{equation}\label{eq:A}
A_1=
\begin{cases}
7/5 \quad& \text{for $\phi_+ \in(0,\pi)$ and $\phi_-\in(0,\pi)$},\\
4/5 \quad& \text{for $\phi_+=0$ and $\phi_- \in(0,\pi)$},\\
4/5 \quad& \text{for $\phi_+=\phi_-=0$ and $k_+=k_-$},\\
1/5 \quad& \text{for $\phi_+=\phi_-=0$ and $k_+\neq k_-$},
\end{cases}
\end{equation}
and
\begin{equation}\label{eq:B}
B_1=
\begin{cases}
\log\left|\frac{10\sin(3\beta-3\chi_1)+10\sin(3\gamma_1)}{\cos\gamma_1}\right| &\text{for $\phi_+ \in(0,\pi)$ and $\phi_-\in(0,\pi)$},\\
\log\left|\frac{\sqrt{30}\cos(3\gamma_1)}{\cos\gamma_1}\right| & \text{for $\phi_+=0$ and $\phi_- \in(0,\pi)$},\\
2\log 3 \quad& \text{for $\phi_+=\phi_-=0$ and $k_+=k_-$},\\
\frac{1}{2}\log 3& \text{for $\phi_+=\phi_-=0$ and $k_+\neq k_-$},
\end{cases}
\end{equation}
for $\phi_\pm\in[0,\pi)$ and $k_\pm\in\{-1,0,1\}$ defined such that the effective-boundary defects are
$a^{-1}(z)=\hat{b}_\pm=e^{-\im\chi_1}e^{-2\im(\phi_\pm+\pi k_\pm)/3}=\pm e^{-\im(\beta\pm\gamma_1)}$, with  periods $\Upsilon\sim -\beta/\log[h(1)d]$ and  $\Gamma_1\sim 4\pi\sin\gamma_1/h(1)$.

\subsection{Secondary triangle}

A similar argument can be applied to the director field local to the secondary triangle, \ie~\eqref{eq:asym_secondarydirector}. It follows  that  
\begin{equation}\label{eq:E2_asymp}
    \hatE_2(\Gamma_2,\Upsilon)\sim \pi A_2\log\left|\frac{ w }{10h(1)}\right|+\pi B_2+\frac{\pi \beta^2}{\log\left|h(1)d\right|},
\end{equation}
as $w\to\infty$,  with the coefficients
\begin{equation}\label{eq:A2}
A_2=
\begin{cases}
7/5 \quad& \text{for $\Phi_+ \in(0,\pi)$ and $\Phi_-\in(0,\pi)$},\\
4/5 \quad& \text{for $\Phi_+=0$ and $\Phi_- \in(0,\pi)$},\\
4/5 \quad& \text{for $\Phi_+=\Phi_-=0$ and $K_+=K_-$},\\
1/5 \quad& \text{for $\Phi_+=\Phi_-=0$ and $K_+\neq K_-$},
\end{cases}
\end{equation}
and
\begin{equation}\label{eq:B2}
B_2=
\begin{cases}
\log\left|\frac{10\sin(3\beta+3\chi_2)+10\sin(3\gamma_1)}{\cos\gamma_2}\right|  &\text{for $\Phi_+ \in(0,\pi)$ and $\Phi_-\in(0,\pi)$},\\
\log\left|\frac{\sqrt{30}\cos(3\gamma_2)}{\cos\gamma_2}\right| & \text{for $\Phi_+=0$ and $\Phi_- \in(0,\pi)$},\\
2\log 3 \quad& \text{for $\Phi_+=\Phi_-=0$ and $K_+=K_-$},\\
\frac{1}{2}\log 3& \text{for $\Phi_+=\Phi_-=0$ and $K_+\neq K_-$},
\end{cases}
\end{equation}
for $\Phi_\pm\in[0,\pi)$ and $K_\pm\in\{-1,0,1\}$ defined such that the effective-boundary defects are $A^{-1}(z)=\hat{B}_\pm=e^{-\im\chi_2}e^{-2\im(\Phi_\pm+\pi K_\pm)/3}=\pm e^{\im(\beta\mp\gamma_2)}$, with periods  $\Upsilon\sim -\beta/\log[h(1)d]$ and  $\Gamma_2\sim 4\pi\sin\gamma_2/h(1)$.

\subsection{Energy minimum}
Combining \eqref{eq:E1_asymp} and \eqref{eq:E2_asymp} yields the net energy
\begin{equation}
    \hatE = \hatE_1(\Gamma_1,\Upsilon)+\hatE_2(\Gamma_2,\Upsilon)\sim \pi (A_1+A_2)\log\left|\frac{ w }{10h(1)}\right|+\pi (B_1+B_2)+\frac{2\pi \beta^2}{\log\left|h(1)d\right|},
\end{equation}
as $w,d\to\infty$. Since $\log w\to\infty$ as  $w\to\infty$, the net energy is minimized when $A_1+A_2$ is minimized, implying  that all four effective-boundary defects are located at distinct corners. However, this can only be achieved if the triangles are oriented such that  $\chi_1=\chi_2\mod{\pi/3}$. In general, only three out of the four effective-boundary defects can be located at corners as there are only three unknown periods. The various combinations of three corners correspond to local energy minimum, with $\hatE\sim \pi \log w$ as $w\to\infty$. By comparing these combinations, one finds that a global minimum occurs when  $\gamma_1$, $\gamma_2$, and $\beta$ take their smallest  values. This  occurs when the defects are at the three corners  closest to the horizontal axes, yielding the periods in \eqref{eq:triangle_asym_period1} and \eqref{eq:triangle_asym_period2}, which were found in the main text.

\bibliographystyle{siamplain}
\bibliography{references}

\end{document}

%% file: ex_shared.tex
\usepackage{lipsum}
\usepackage{amsfonts}
\usepackage{mathtools,amssymb}
\usepackage{pifont}
\usepackage{bm}
\usepackage{graphicx}
\usepackage{epstopdf}
\usepackage{algorithmic}
\ifpdf
  \DeclareGraphicsExtensions{.eps,.pdf,.png,.jpg}
\else
  \DeclareGraphicsExtensions{.eps}
\fi

\newcommand{\im}{i}
\newcommand{\e}{e}
\newcommand*\de{\mathop{}\!\mathrm{d}}

\newcommand{\Oh}{\mathcal{O}}

\renewcommand*{\Re}{\operatorname{Re}}
\renewcommand*{\Im}{\operatorname{Im}}
\DeclareMathOperator{\sgn}{sgn}

\newcommand{\myeqref}[2]{(\hyperref[#1]{\ref*{#1}#2})}
\newcommand{\myref}[2]{\hyperref[#1]{\ref*{#1}(#2)}}
\newcommand{\myrefnb}[2]{\hyperref[#1]{\ref*{#1}#2}}
\newcommand{\subtag}[1]{\tag{\theparentequation #1}}

\newcommand*{\ie}{\emph{i.e.}}
\newcommand*{\eg}{\emph{e.g.}}

\newcommand{\w}{w}
\newcommand{\hatE}{\hat{\mathcal{E}}}
\newcommand{\hattau}{\hat{t}}

\newcommand{\hatF}{\hat{F}}
\newcommand{\hatT}{\hat{T}}
\newcommand{\n}{\hat{\nu}}
\newcommand{\s}{\hat{s}}

\newcommand{\ind}{k}
\newcommand{\Di}{D_\ind}
\newcommand{\phii}{\phi_\ind}
\newcommand{\Wi}{W_\ind}
\newcommand{\wi}{w_\ind}
\newcommand{\Ui}{\Upsilon_\ind}
\newcommand{\Mi}{M_\ind}
\newcommand{\ww}{\eta}

\newsiamremark{remark}{Remark}
\newsiamremark{hypothesis}{Hypothesis}
\crefname{hypothesis}{Hypothesis}{Hypotheses}
\newsiamthm{claim}{Claim}

\headers{Solutions for elastic interactions in a nematic liquid crystal}{Thomas G.~J.~Chandler and Saverio E.~Spagnolie}

\title{Exact and approximate solutions for elastic interactions in a nematic liquid crystal\thanks{Submitted to the editors July 24, 2024. \funding{Funding from the NSF (DMR-2003807) is gratefully acknowledged}}}

\author{Thomas G.~J.~Chandler\thanks{Department of Mathematics, University of Wisconsin--Madison, Madison, WI
  (\email{tgchandler@wisc.edu}).}
\and Saverio E.~Spagnolie\footnotemark[2]}

\usepackage{amsopn}

\makeatletter
\newcommand*{\addFileDependency}[1]{% argument=file name and extension
  \typeout{(#1)}% latexmk will find this if $recorder=0 (however, in that case, it will ignore #1 if it is a .aux or .pdf file etc and it exists! if it doesn't exist, it will appear in the list of dependents regardless)
  \@addtofilelist{#1}% if you want it to appear in \listfiles, not really necessary and latexmk doesn't use this
  \IfFileExists{#1}{}{\typeout{No file #1.}}% latexmk will find this message if #1 doesn't exist (yet)
}
\makeatother